\documentclass[12pt]{article}

\usepackage[T1]{fontenc}
\usepackage{microtype}
\usepackage{times}
\usepackage{graphicx}
\usepackage{todonotes}
\PassOptionsToPackage{hyphens}{url}\usepackage[hidelinks]{hyperref}
\usepackage[natbibapa]{apacite}
\usepackage[newfloat,draft]{minted}
\usepackage{footmisc}
\usepackage{multirow}
\usepackage{booktabs}
\usepackage{tabularx}
\usepackage{csquotes}
\usepackage{orcidlink}

\definecolor{mygreen}{RGB}{221,232,203}
\definecolor{myorange}{RGB}{254,217,166}
\definecolor{mypink}{RGB}{255,215,215}
\definecolor{mybrown}{RGB}{229,216,189}
\definecolor{mybeige}{RGB}{255,219,182}
\definecolor{myblue}{RGB}{218,229,241}
\definecolor{myviolet}{RGB}{224,194,205}
\definecolor{myadmiral}{RGB}{189,201,225}
\definecolor{myred}{RGB}{246,161,117}
\definecolor{light}{RGB}{239,101,72}
\definecolor{error}{RGB}{215,48,31}
\definecolor{error2}{RGB}{153,0,0}
\providecommand{\keywords}[1]
{
  \small	
  \textbf{\textit{Keywords---}} #1
}

\newenvironment{sciabstract}{%
\begin{quote} \bf}
{\end{quote}}

\newcommand\authorbox[2][]{\tikz[overlay]\node[fill=myorange,ultra thick,inner sep=2pt, anchor=text, rectangle,#1] {#2};\phantom{#2}}
\newcommand\namebox[2][]{\tikz[overlay]\node[fill=myadmiral,ultra thick,inner sep=2pt, anchor=text, rectangle,#1] {#2};\phantom{#2}}
\newcommand\descbox[2][]{\tikz[overlay]\node[fill=myred,ultra thick,inner sep=2pt, anchor=text, rectangle,#1] {#2};\phantom{#2}}
\newcommand\verbox[2][]{\tikz[overlay]\node[fill=mygreen,ultra thick,inner sep=2pt, anchor=text, rectangle,#1] {#2};\phantom{#2}}
\newcommand\typebox[2][]{\tikz[overlay]\node[fill=mybrown,ultra thick,inner sep=2pt, anchor=text, rectangle,#1] {#2};\phantom{#2}}
\newcommand\doabox[2][]{\tikz[overlay]\node[fill=myviolet,ultra thick,inner sep=2pt, anchor=text, rectangle,#1] {#2};\phantom{#2}}
\newcommand\urlbox[2][]{\tikz[overlay]\node[fill=myblue,ultra thick,inner sep=2pt, anchor=text, rectangle,#1] {#2};\phantom{#2}}
\newcommand\yearbox[2][]{\tikz[overlay]\node[fill=mypink,ultra thick,inner sep=2pt, anchor=text, rectangle,#1] {#2};\phantom{#2}}

\newcommand\unstructuredbox[2][]{\tikz[overlay]\node[draw=error,dashed,ultra thick,inner sep=0pt, anchor=text, rectangle,#1] {#2};\phantom{#2}}

\newcommand\incompletebox[2][]{\tikz[overlay]\node[draw=error,dotted,ultra thick,inner sep=0pt, anchor=text, rectangle,#1] {#2};\phantom{#2}}

\newcommand\misrepbox[2][]{\tikz[overlay]\node[draw=error,dotted,ultra thick,inner sep=0pt, anchor=text, rectangle,#1] {#2};\phantom{#2}}

\newcommand\wrongbox[2][]{\tikz[overlay]\node[draw=error,ultra thick,inner sep=0pt, anchor=text, rectangle,#1] {#2};\phantom{#2}}

\title{A multi-level analysis of data quality for formal software citation}

\author
{David Schindler$^{1}$\orcidlink{0000-0003-4203-8851}, Tazin Hossain$^{1}$\orcidlink{0009-0007-7011-6796}, Sascha Spors$^{1}$\orcidlink{0000-0001-7225-9992}, Frank Krüger$^{1,2,3\ast}$\orcidlink{0000-0002-7925-3363}\\
\\
\normalsize{$^{1}$Institute of Communications Engineering, University of Rostock, Rostock, Germany}\\
\normalsize{$^{2}$Faculty of Engineering, University of Applied Sciences, Wismar, Germany}\\
\normalsize{$^{3}$Department Knowledge, Culture \& Transformation, University of Rostock, Rostock, Germany}\\
\\
\normalsize{$^\ast$To whom correspondence should be addressed; E-mail:  frank.krueger@hs-wismar.de.}
}

\begin{document}
\maketitle              
\begin{sciabstract}
Software is a central part of modern science, and knowledge of its use is crucial for the scientific community with respect to reproducibility and attribution of its developers.
Several studies have investigated in-text mentions of software and its quality, while the quality of formal software citations has only been analyzed superficially. 
This study performs an in-depth evaluation of formal software citation based on a set of manually annotated software references.
It examines which resources are cited for software usage, to what extend they allow proper identification of software and its specific version, how this information is made available by scientific publishers, and how well it is represented in large-scale bibliographic databases.
The results show that software articles are the most cited resource for software, while direct software citations are better suited for identification of software versions. 
Moreover, we found current practices by both, publishers and bibliographic databases, to be unsuited to represent these direct software citations, hindering large-scale analyses such as assessing software impact.
We argue that current practices for representing software citations---the recommended way to cite software by current citation standards---stand in the way of their adaption by the scientific community, and urge providers of bibliographic data to explicitly model scientific software. 
\end{sciabstract}



%
%
\section{Introduction}
\label{sec:intro}

Software is an important part of modern science and contributes to the provenance of research results.
From the microscopic perspective, the identification of the particular software, and its specific version, that was used for a respective study is important to allow the reproduction of the results.
The macroscopic perspective to provenance enables large scale analyses of software impact similar to impact factors for scholarly publications and thus allows to provide credit to the developers and funders of the software and to analyse patterns of software usage across research domains.
In general, software is considered one of the main pillars of science---besides articles and data---as it contains the logic of data transformation~\citep{di2020curated}.  
Therefore, it is advocated that its contribution to research should be indicated and formally cited~\citep{smith2016software,Katz2021}.   
The value of software is also recognized by different stakeholders in the scientific community, with journal policies requiring indication of software usage, and funders requiring researchers to make developed software and source code available as research results.
Moreover, researchers themselves have taken up the role of software developers, with 84\% reporting that developing software is essential for their research~\citep{Goble2014}, while there is also an increasing need for funding allocated for the development of research software~\citep{hong2016we}.  

Software development cannot be properly recognized without being included in measures of impact \citep{wright2023open}.
In science, such measures rely on bibliographic analysis. 
With respect to research software, however, citation analyses are currently limited to in-text software mentions, and have been performed by automatic identification of software mentions in the full-text documents~\citep{Schindler2022TheRO,istrate2022large,Du2022}.
On the one hand, this is due to historic reasons because the high impact of software on research has only recently been recognized by the scientific community and formal software citation has only recently been advocated. 
On the other hand, it is not clear whether the current citation practices and the infrastructure for citation analysis are suited to represent software, and whether they could be utilized for such analyses. 
Particularly, since software should not be cited through a proxy, such as a software article, but directly to allow its proper identification. 
This is crucial since aspects of software citation differ from the citation of articles, with the specific requirements of proper software citation defined by citation guidelines~\citep{Katz2021}.
Specifically, versioning---essential for provenance and reproducibility---is not considered in article citation. 
Hence, technical updates might be necessary to create suited and machine-readable software representations~\citep{stall2023journal}.

Bibliometric and citation analyses are performed with the aid of providers of bibliometric data and are based on the structured data they provide. 
Semantic Scholar~\citep{kinney2023semantic} or Crossref~\citep{hendricks2020crossref}, for instance, provide powerful application programming interfaces (API) to access the already pre-processed data about millions of articles.
Such infrastructure should also build the basis to integrate software in bibliometric analyses.
However, beside the structure of the provided data, such analyses also heavily depend on the data quality~\citep{Haustein2014}. 
With respect to software citations (and citations in general) this includes different stages of data collection and processing beginning with the authors of scholarly publications who use software and provide all information necessary to identify the particular software and the actual code base\footnote{This term refers to the specific development state of a software, typically indicated by a version.}, ranging to publishers to provide structured data, and bibliographic databases that collect and provide the data, which is later used for scientometric studies.
According to \cite{Batini2009}, measurements of data quality consist of different dimensions including accuracy and completeness, where the first describes the correctness of the provided data and the second describes whether all necessary information is covered.
Regarding software citations, completeness can be interpreted with respect to the identification of the software and the particular code base.
Accuracy, in contrast, describes to what extend processed data, i.e., provided by databases, reflect the original content as provided by the authors. 

In this article, we analyse the data quality of software citations across the entire data lifecycle, beginning with references as initially provided by the authors, provided by publishers, and finally as provided by two major databases for bibliographic data.
All analyses we perform are based on a high-quality, manually annotated dataset, established in the scope of this work by extending the existing gold standard corpus {\sc SoMeSci} of software mentions in scholarly publications. 
We first examine what exactly formal software citations refer to and investigate the completeness of such citations with respect to (1) the particular software, (2) software creator attribution, and (3) the particular software version.
Finally, we evaluate whether the bibliometric databases Semantic Scholar and Crossref can actually be used to estimate the impact a software might have.

The results of our analyses show that formal software citations most frequently refer to software articles, illustrating the importance of such as a surrogate citation target of the software.
While this typically does not help to identify the particularly employed software version, it certainly allows the identification of the software itself and provides credit for its development. 
When using direct software citation, we find that only about 2/3 of them allow the identification of the actual software code base.
With respect to bibliometric data providers, we find that significant parts of direct software citations are not represented or contain errors.
We presume that algorithms for matching references to scholarly articles often produce wrong results when applied to direct software references and, therefore, conclude that such databases are currently not suited for large scale analyses of software citation patterns.
Furthermore, our work shows how different stakeholders in science---especially authors, software developers, and providers of bibliographic data---can contribute to improve the traceability and identification of software in scientific literature. 

\section{Related Work}
\label{sec:related_work}

As outlined in Section~\ref{sec:intro}, software is ubiquitous in data driven science, and knowledge of its use essential for the scientific community. 
Recent work, has found that software is either mentioned informally within the full-text document of scientific articles, and formally with a bibliographic reference, with the first practice being more common~\citep{Howison:2016,SoMeSci,Schindler2022TheRO,Du2022}. 
The analyses of informal mentions has been subject of multiple investigations, either performing high-quality manual analyses on small corpora~\citep{Howison:2016,Du2021,nangia.katz:2017,SoMeSci} or automatic large-scale analyses~\citep{pan2015assessing,duck2016survey,own_2,Schindler2022TheRO}. 
The reported results often strongly vary due to different underlying data and chosen approach. 
The scientific domain, for instance, has a strong influence on software usage, with numbers between 0.2 software mentioned per article reported in Economics up to 30.8 in Bioinformatics.
Some work has further included analyses on how often formal citations are provided together with informal mentions, with results varying due to approach and underlying data. 
\cite{Howison:2016} report that 44\% of informal mentions include a formal citation, while \cite{SoMeSci} report 16\%, 24.8\% are found in the data of \cite{Schindler2022TheRO}, and \cite{Du2022} report 18\%.
Only the work of \cite{Howison:2016} and \cite{Du2022} further investigate the formal citations themselves.
They distinguish the cited resource behind formal citations and report that respective 84\% and 89\% of citations refer to articles. 
\cite{Howison:2016} further identified 5\% of citations referring to software manuals and 11\% to software directly, while \cite{Du2022} report 8\% referring to software directly.  
In this work, further analyses on resource types for software citations are performed to show validity of the data. 

Identification, credit, and provenance are three central aspects for software citation \citep{Soito:2016,smith2016software}.
The identification of a software is considered possible when the provided meta-data allows to uniquely determine the used software. 
Software names are, in general, insufficient for this purpose because they have been shown to be ambiguous~\citep{duck2015ambiguity,Schindler2022TheRO} and can potentially refer to legacy software that is no longer findable by name. 
Therefore, the use of persistent unique identifiers is advocated for software citations~\citep{Soito:2016,smith2016software,Katz2021}.  
Credit and attribution for the development of software is important for multiple stake-holders who have an interest in assessing the impact of a software, including software developers and research funders.
Software has not consistently been treated as citable resource by the scientific community~\citep{bouquin2020credit}, which made it hard to assess its impact and to provide proper credit for its costly development~\citep{mayernik2017assessing}. 
In general, proper attribution of a software developer can be challenging when multiple people or instances with different contributions are involved~\citep{katz2015transitive}, or even impossible for open source projects.  
The use of software is part of a research's provenance, therefore, not only the software but also its specific development state---referred to as code base in this work and usually indicated by a version---needs to be uniquely identifiable by the meta-data provided with a software citation because most software is under constant development and changing in its range of functions and behavior~\citep{smith2016software,Katz2021}. 
In general, the development state can either be uniquely identified by version numbers assigned by the developer or by a release date corresponding to a version~\citep{Katz2021}. 
The completeness of informal software mentions with respect to identification, credit, and provenance has received some attention in the existing literature~\citep{Howison:2016,Du2021,SoMeSci,Schindler2022TheRO,Du2022}, which found that current mention practices often lack information. 
Again, only the work of \cite{Howison:2016} and \cite{Du2022} takes formal references into account in this context, by including the information in formal references when determining the overall completeness of software citations.
Moreover, \cite{Du2022} explicitly provide analyses of formal citations and report that 35\% include a version and 78\% identify software developers. 
In this work, we perform further systematic analyses of the completeness of formal software citation in terms of identification, credit, and provenance, to gain a better understanding of software citation practices in scientific literature. 

Large scale scientometric analyses are commonly performed with bibliographic databases~\citep{Napolitano2021,dion2018gendered,peroni2020practice,cho2018link}, as they provide the necessary structured meta-data for formal citations. 
Software could be included in such analyses if its formal citations are represented within those databases and semantically correctly structured. 
A semantic representation is important because metadata that is not correctly structured can become useless for downstream tasks.
Proper representation of the software name and version, for instance, is crucial for tracking of software usage, and are necessary for the disambiguation of citation targets by data providers themselves. 
As described, \citet{stall2023journal} argue that updates to existing infrastructure might be necessary for this purpose. 
We analyze how well software references are represented within the state of the art bibliographic databases Semantic Scholar\footnote{\url{https://www.semanticscholar.org/}} and Crossref\footnote{\url{https://www.crossref.org/}} to asses how well they can be used as scientometric resources with respect to the analyses of scientific software usage. 
Semantic Scholar is a discovery service for scientific literature~\citep{wade2022semantic} developed and maintained by the  Allen Institute for Artificial Intelligence (AI2).
The service is based on the Semantic Scholar Academic Graph (S2AG) containing meta-data of scientific publications for 205M publications and 2.5B citation edges~\citep{wade2022semantic}.
A major aspect of Semantic Scholar is to integrate machine learning methods to enhance data quality and search. 
They did, for instance, develop a system for publication deduplication named S2APLER and perform citation linking based on fuzzy text-matching heuristics~\citep{kinney2023semantic}. 
Data is provided free and open by Semantic Scholar and can be accessed via API.
Semantic Scholar is widely used as a search mechanism for academic publications, while the underlying knowledge graph also enables scientometric analyses~\citep{Napolitano2021}. 

Crossref is the result of a joint effort by an association of publishers~\citep{lammey2015crossref} with over 17,000 members as of June 2023, with the goal to improve linking between publications made by heterogeneous publishers.
The main application of Crossref is a database of meta-data for scholarly articles and professional materials that enables unique identification of covered resources by incorporating and introducing persistent identifiers. 
The corresponding bibliographic information is integrated into the database by publishers with central quality control by Crossref.
Moreover, meta-data is enriched by Crossref, mainly by adding citation links, but also through adding further information such as funder registry information or journal classification codes~\citep{hendricks2020crossref}. 
Crossref makes the database openly available without any license restriction through an API and, therefore, enables scientometric analyses.
The resource is widely used for this purpose, for instance, for citation analyses by \citet{dion2018gendered,peroni2020practice} and citation network analyses by \citet{cho2018link}.

\section{Analyses}
\label{sec:analyses}
The goal of this study is to investigate the data quality of formal software citation in science based on manually annotated, high quality data. 
This sections outlines the four main research questions and the analyses employed to investigate them: 
\begin{itemize}
    \item What types of resources are referenced by formal software citations?
    \item Is software formally cited without being mentioned in the full-text document?
    \item Do formal software references provide all necessary information to identify software, developer, and the used code base?
    \item How well are formal software references represented in bibliographic databases?
\end{itemize}
Answering these questions, allows to assess the state of formal software citation and to reveal potential short-comings in current practices, which can be used to formulate recommendations on how to improve the quality of formal software citation and software traceability in scientific literature. 

\subsection{Citations Resource Types}
\label{sec:ana_formal_citation_type}
The first analysis investigates what type of resource is referenced by the bibliographic entry associated to in-text software mentions in scientific publications.
Different resources can be referenced within these bibliographic entries because different software citation practices exist in the scientific community. 
Not all of these practices are suited for formal software citation because they might not provide all necessary information to identify the software. 
Analyzing them, therefore, allows to assess current practices and can reveal shortcomings.

The resource type is analyzed by determining the distribution of resource types in the dataset introduced in Section~\ref{sec:data}.
Furthermore, the resource type is analyzed with respect to the meta-data provided in the context of the corresponding informal in-text software mention because authors using unsuited citation types that do not provide all required meta-data to identify a software could systematically add the missing information in the full-text document. 
The relevant resource types were defined based on the previous work of~\citet{Howison:2016}, who distinguished between citations to publications, user manuals, and project names or websites, and further extended based on observations made during data annotation.

\paragraph*{Direct Software Citations} describe the cited software itself and are the recommended way to cite software by recently established software citation guidelines~\citep{Katz2021}. 
An example for a \emph{Direct Software Citation}, from here on referred to as \emph{Direct Citation}, is given in Listings~\ref{list:software_direct_txt} and~\ref{list:software_direct_xml}, corresponding to the bibliographic entry as available in the PDF publication and in the Journal Article Tag Suite (JATS) XML. 
\begin{listing}[tb]
    \includegraphics[width=\textwidth]{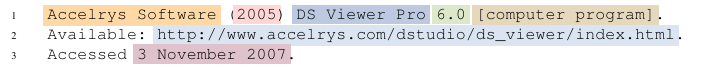}
    \caption{Example for a \emph{Direct Citation} from the dataset described in Section~\ref{sec:data} [ID: PMC2134966, \cite{del2007modular}]. The content of the bibliography entry is provided as in the original PDF publication. Information closer identifying the software is highlighted: \authorbox{Developer}, \yearbox{Year}, \namebox{Name}, \verbox{Version}, \typebox{Type of Citation}, \urlbox{URL}, \doabox{Date of Access}.}
    \label{list:software_direct_txt}
\end{listing}
\begin{listing}[tb]
    \includegraphics[width=\textwidth]{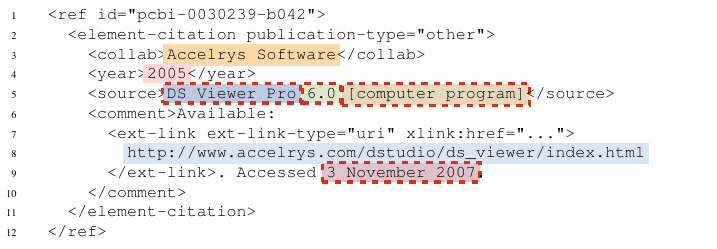}
    \caption{Corresponding JATS entry for the \emph{Direct Citation} in Listing~\ref{list:software_direct_txt}, highlighting the same information.
    Additionally, meta-data that is not structured with a corresponding label is \unstructuredbox[inner sep=2pt]{highlighted}.}
    \label{list:software_direct_xml} 
\end{listing}
Properly executed \emph{Direct Citations} capture all meta-data of software that is required for a unique identification of the software, its developer, and the exact code base. 
In practice, not all required information may be present and the references themselves can be arbitrary structured as there is no commonly agreed citation style for direct software citations, which is further investigated in Section~\ref{sec:ana_cit_comp}.

\paragraph{Software Articles} are scientific articles describing a scientific software that are published by developers of scientific software, and cited in place of the software~\citep{Howison:2016}.
An example is provided in Listing~\ref{list:software_article}.
\begin{listing}[tb]
    \includegraphics[width=\textwidth]{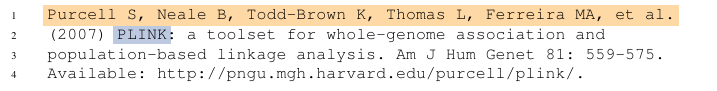}
    \caption{Example for a \emph{Software Article} reference from the  dataset established in Section~\ref{sec:data} [ID: PMC2987837, \cite{paternoster2010genome}]. All information describing the software highlighted with \authorbox{Developer} and \namebox{Name}. Note that a publication year is provided, but it describes the original publication date of the article and does not provide any information about the release of the software.}
    \label{list:software_article}
\end{listing} 
The practice is common as it allows developers of scientific software to receive scientific attribution for the costly development of the research software, and, historically, the publication of a software has been considered a weaker contribution as the publication of an article~\cite{hafer2009assessing}.
Software articles are among the highest cited scientific papers~\citep{van2014top} and specific journals for publishing software articles have been established, e.g., \emph{The Journal of Open Research Software} or \emph{Source Code for Biology and Medicine}.
Software articles are cited the same way as other scientific articles without software specific information. 
Therefore, information identifying the code basis such as the version or release date is generally missing from this citation type. 

\paragraph{Software Manuals} are textual instructions for using a software, and, particularly for commercial software, they are often the closest textual document associated with a software.
It is an established practice to cite such manuals instead of the software itself with an example given in Listing~\ref{list:software_manual}.
\begin{listing}[tb]
    \includegraphics[width=\textwidth]{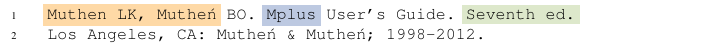}
    \caption{Example of a \emph{Software Manual} reference from the dataset established in Section~\ref{sec:data} [ID: PMC5597179, \cite{maraz2017pathological}]. Information identifying the software, its developer, and the used code based are highlighted: \authorbox{Developer}, \namebox{Name}, \verbox{Version}. Note that information on the version or publication date of a manual can in some instances be used to identify the software code base if the manual is updated with every release of the software and authors cite it appropriately.  }
    \label{list:software_manual}
\end{listing}
Corresponding references are formatted for citing a text source with information typically provided for online source such as an URL and date of access.
Same as software articles, this citation type omits crucial information closer describing the corresponding software.

\paragraph{Websites} associated with a software are sometimes cited instead of a software.
The corresponding references are structured as typical online resources, potentially providing the date of access. 
An example is provided in Listing~\ref{list:software_website}.
\begin{listing}[tb]
    \includegraphics[width=\textwidth]{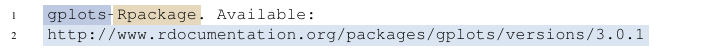}
    \caption{Example of a \emph{Software Website} reference from the dataset established in Section~\ref{sec:data} [ID: PMC5070780, \cite{malvisi2016responses}].
    Information identifying the software, its developer, and the used code based are highlighted: \namebox{Name}, \typebox{Type of Citation}, \urlbox{URL}.}
    \label{list:software_website}
\end{listing}
Same as for other styles, relevant information specifying the software itself is missing. 
As \emph{Direct Citations} often include URLs, it is important to distinguish them from \emph{Websites}. 
Here, all cases where additional information about the software is provided (except name, URL, and date of access) are considered as \emph{Direct Citations}.

\paragraph{Other} citations describe the rare instances in which no verifiable resource is described by a reference.
Those cases are present in practice and are likely to result from faulty automatic citation recommendations or author errors. 

\subsection{Formal Software Reference without in-text Software Mentions}
The second part of the analysis investigates if software is formally referenced even if its not mentioned within a article's full-text document. 
In theory, it is possible that authors formally cite software but do not state the name of a software, for instance, if they replace the software name with a generic term such as ``source code''. 
\citet{Howison:2016} report that generic terms only make up 1\% of overall software mentions, but there could be further reasons software is formally cited but not mentioned. 
Since this aspect has---to the best of our knowledge---not yet been analyzed, we investigate if this practice exists and include the resulting set of references in the analyses on completeness.
However, we only consider \emph{Direct Citations}, \emph{Websites} and \emph{Manuals} because it is not feasible to annotate \emph{software articles} as explained in Section~\ref{sec:anno}.
Analyzing this trend is highly relevant to formal software citations because it allows a better assessement of its importance for software traceability, since software usage is only identifiable through formal citations  in the described cases, which is not considered by current methods analyzing software in scientific publications. 

\subsection{Direct Citation Completeness}
\label{sec:ana_cit_comp}

The third part of the analysis investigates the completeness of \emph{Direct Citations}, \emph{Manuals}, \emph{Websites}\footnote{\emph{Direct Citations} and \emph{Websites} are summarized under the assumption that \emph{Website} are incomplete \emph{Direct Citations}.} in terms of meta-data describing the software as provided by the authors of the scientific publication, which have the responsibility of providing complete information to identify the software.
In general, \emph{Direct Citations} are the recommended practice for software citation~\citep{Katz2021} because they allow unique identification of software, developer, and the exact code base. 
However, it has not yet been analyzed what meta-data is actually provided in practice aside from the version and developer~\citep{Du2022}.
Therefore, the completeness is analyzed in terms of: \emph{Name}, \emph{Creator}, \emph{Identifier}, \emph{Archive}, \emph{URL}, \emph{Release Date} (exact or only year), \emph{Version}, \emph{Date of Access}, \emph{Type of Citation}, and \emph{Description}, by analyzing the number of cases where the information was provided. 
An \emph{Identifier} is defined as a specific unique identifier for a software, e.g., an RRID; an \emph{Archive} is assumed to be a persistent link to a repository where a software is published; a \emph{URL} is any other link that is provided; and the \emph{Type of Citation} is usually provided to identify the type of source that was cited, e.g., ``[Source Code]'' or ``[Software]'' as shown in Listings~\ref{list:software_direct_txt} and~\ref{list:software_website}.
These specific metadata were selected as they are the recommended information to be provided for software citations to allow proper identification of the software, where \emph{Identifier}, \emph{Archive}, or \emph{URL} can be applied in the given priority dependent on how software was published, while a \emph{Date of Access} and \emph{Version} both allow a unique identification of the exact software development state, and \emph{Type of Citation} as well as \emph{Description} are considered as optional information~\citep{Katz2021}.

Further, it is analyzed whether the referenced software is identifiable, whether its developer can be attributed, and whether the specific code base can be determined.
Software is considered as \textbf{identifiable} if either $\bullet$~\emph{Identifier}, $\bullet$~\emph{Archive}, or $\bullet$~\emph{Name} and \emph{Creator} are provided.
Furthermore, references providing \emph{URLs} are considered as conditionally identifiable because URLs are not persistent and commonly become invalid over time, with the effect of link rot regarding scientific data having been shown in prior work by \cite{lakic2023link}.
Moreover, they can link to different resources associated to a software, e.g., the download website, reference manuals, or the creator.
The information for proper \textbf{attribution} is considered given if the software is identifiable and its \emph{Creator} is stated.
The \textbf{code base} refers to the exact software development state, and is defined as identifiable when either $\bullet$~\emph{Release Date} or $\bullet$~\emph{Version} number is given. 
A \emph{Date of Access} is considered as conditionally identifiable under the assumption that the newest available release was used. 
The release year is considered as insufficient as multiple minor versions or even more than one major version can exist in the same year.
Lastly, the overall completeness for software citations is analyzed if a software has a corresponding informal in-text software mention because authors could choose to provide part of the meta-data within the full-text document and part of the meta-data in the formal citation. 

This analysis is of central importance since authors have the main responsibility in providing complete information in their software citations which provides the basis for representation by publishers and literature databases. 
It can reveal how well software is formally cited, and builds the basis to formulate recommendations for potential improvements. 
These can serve as a basis for funders or journals to update their policies and require proper attribution of software usage by formal citation as recommended.

\subsection{Database Accuracy}
\label{sec:ana_data_rep}

The fourth analysis investigates how well \emph{Direct Citations} are represented by publishers and especially within scientific bibliographic databases. 
There are several studies analyzing informal software mentions in scientific literature (see Section~\ref{sec:related_work}), while no large scale analyses of formal references has been performed.
To implement such bibliometric analyses scientists typically employ large scale bibliographic databases, while the information of the databases is based on the information provided by scientific publishers, who need to mark up the metadata provided by authors in a suited manner.  
However, it is not clear how \emph{Direct Citations} are represented by both publishers and databases because the structure of software citations strongly varies from other scientific publications that are usually represented by them \citep{stall2023journal}. 
Regarding publishers, it is quantitatively analyzed how software citations are structured from the publishers side to assess the quality of the semantic representation.
With respect to bibliographic databases, it is quantitatively analyzed what information is available within the databases, whether the information is represented in a structured manner that would allows a systematic analysis, and whether the information provided within the database is correct.
For correctness it is considered whether the information from the originally provided reference differs from the information contained in a database, but also whether new information, which can potentially be added by a database, is correct. 
All aspects are examined for all individual references in comparison between the information provided by the publisher and the state of the art bibliographic databases Crossref and Semantic Scholar.

This analyses is essential with respect to formal software citations because it allows to assess whether they can be systematically utilized by the scientific community based on the current infrastructure of publishers and bibliographic databases.
This could, in turn, allow an extension and enhancement of current analyses of software citation. 
On the other hand, the analysis has the potential to gather insights in current short-comings of the representation of formal software citations and provide the basis to update the infrastructure of bibliographic data providers.

\subsection{Confidence Intervals (CI)}
\label{sec:confidence_intervals}
Confidence Intervals are used for reporting of statistical results with the underlying data being either multinomial or binary. 
Multinomial CIs are calculated based on the method proposed by \cite{glaz1999simultaneous,sison1995simultaneous} for calculation of simultaneous CIs. 
For binary variables, CIs are based on the binomial distribution, calculated as commonly known under the term Wald-interval, using an approximation by a normal distribution as proposed by \cite{wallis2013binomial}.

\section{Dataset}
\label{sec:data}

A high-quality dataset to analyze all aspects outlined in Section~\ref{sec:analyses} was established by with quality control at each annotation step. 
The dataset was based on the SoMeSci corpus of informal software mention in scientific publications, which was extended to also cover formal citation.
SoMeSci is a manually annotated dataset with data quality ensured by high Inter Annotator Agreement (IAA) of $\kappa{=} .82$. 
It covers several aspects of software mentions within scientific literature of which we utilize the information on informal software mentions and their associated formal citations.  
The approach of extending SoMeSci was chosen as the performed annotation required considerable manual effort (details described below), which could be systematically reduced based on SoMeSci, without restricting gold standard annotation quality.   

\begin{figure*}[tb]
    \includegraphics[width=\textwidth]{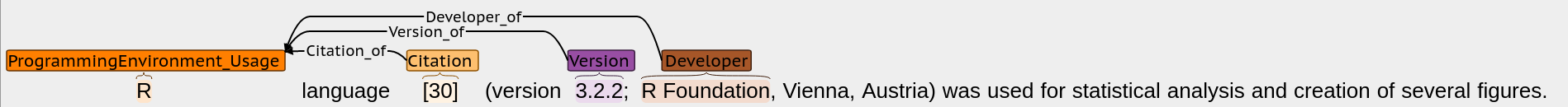}
    \caption{{\sc SoMeSci} annotation example where the software \emph{R} is mentioned in-text and the citation ``[30]'' is associated to the mention. Example from~\cite{SoMeSci}.}
    \label{fig:somesci_ex}
\end{figure*}

Overall, SoMeSci contains 1367 articles in four sets with varying sources and annotation properties: 
\emph{PLoS methods} includes 480 methods sections from articles published by the open source scientific publisher PLoS;
\emph{PLoS sentences} includes selected sentences from 677 articles published by PLoS;
\emph{PubMed fulltext} includes 100 full-text document publications from the PubMed Central Open Access (PMC OA) set;
\emph{Creation sentences} includes selected sentences from 110 articles selected from PLoS and PMC OA that specifically publish software. 
Within those articles, a total of 3756 in-text mentions of software and 591 corresponding in-text citations are contained. 
An example of such an annotation is illustrated in Figure~\ref{fig:somesci_ex} with the corresponding bibliographic entry provided in Listing~\ref{list:xml_ref}. 
\begin{listing}[tb]
    \includegraphics[width=\textwidth]{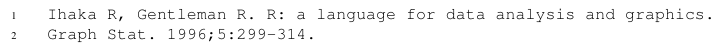}
    \caption{Reference information to the {\sc SoMeSci} annotation \#30 of article \emph{PMC5690316} given in Figure~\ref{fig:somesci_ex}. The reference is an example of a \emph{Software Article}.}
    \label{list:xml_ref}
\end{listing}

\subsection{Annotation}
\label{sec:anno}
\begin{figure*}[tb]
    \centering
    \includegraphics[width=0.9\textwidth]{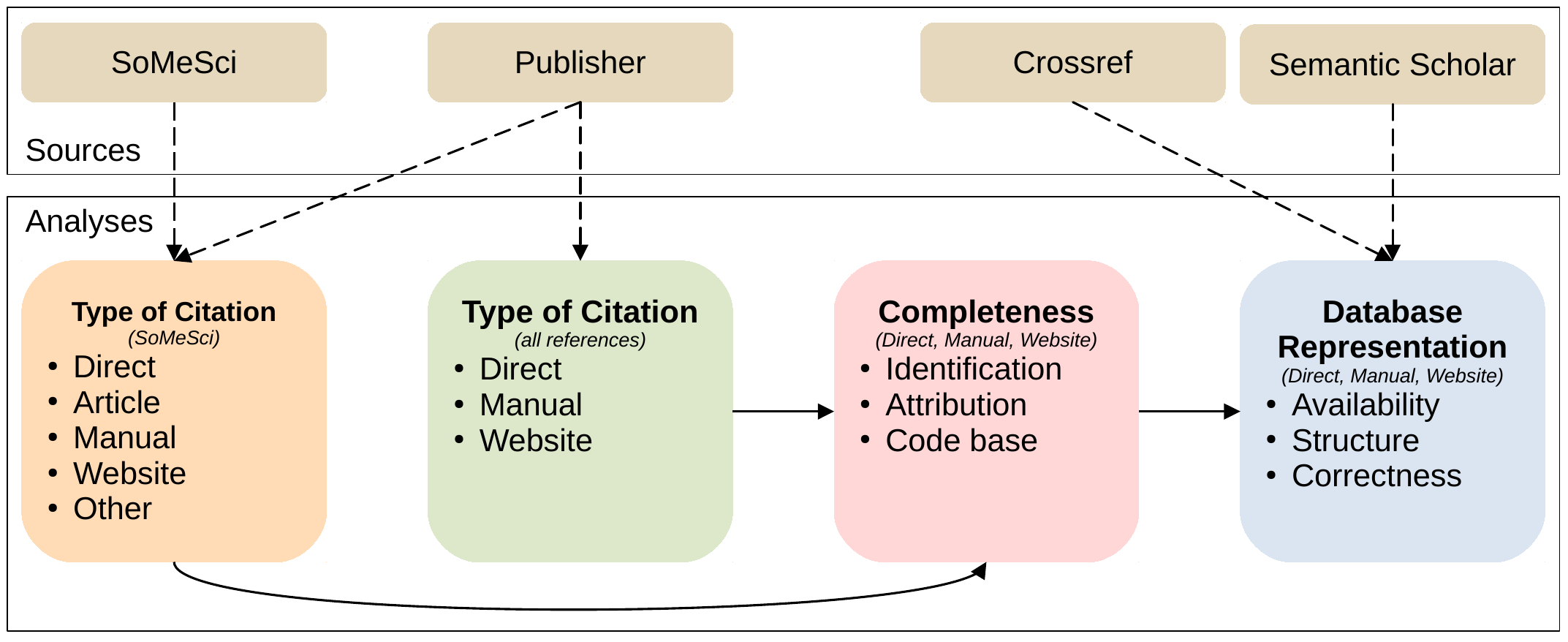}
\caption{Flowchart illustrating the annotation and analyses steps to investigate the research questions outlined in  Section~\ref{sec:analyses}.}
    \label{fig:pipeline}
\end{figure*}

Data was systematically annotated to answer all research questions outlined in Section~\ref{sec:analyses}.
An overview of the data annotation for analyses and the corresponding data sources is given in Figure~\ref{fig:pipeline}.
Since high quality data is required to make reliable statements, data quality was evaluated at every step throughout the annotation process. 
The quality was assessed by calculating the Inter Annotator Agreement (IAA) based on Cohen's $\kappa$~\citep{cohen1960coefficient} to account for chance agreement for the categorical annotation tasks. 
Particularly challenging annotations with insufficient agreement were handled by double annotation and re-annotation of diverging cases.

\paragraph{Software Citation Types} (as outlined in Section~\ref{sec:ana_formal_citation_type}) were manually annotated for all references associated to an in-text software mention in {\sc SoMeSci} (Figure~\ref{fig:pipeline}, left). 
The reference content itself is not contained in {\sc SoMeSci} and was obtained from the publishers in January 22, 2022\footnote{This is when the {\sc SoMeSci} dataset was obtained.} and the reference type was then annotated based upon it.
An initial overlapping annotation of 10\% of data showed an IAA of $\kappa {=} .75$, which can be considered substantial agreement~\citep{landis1977measurement} but was determined as insufficient for analyses. 
Therefore, all samples were annotated by two annotators.
To test the consistency of the annotation we also evaluated the final overall agreement at a value of $\kappa {=} .76$. 
Differences were then discussed and re-annotated to ensure high data quality.  

\paragraph{Formal Software Citation without in-text Mentions}
To identify cases where software is formally referenced without being mentioned in-text (Figure~\ref{fig:pipeline}, middle-left), all 28,903 bibliographic references in the combined 579 of \emph{PLoS Methods} and \emph{PubMed Fulltext} sets of {\sc SoMeSci} were annotated for citation type. 
The remaining articles were not included to keep the number of references and the corresponding annotation cost at a feasible level. 
The annotation was further implemented in two steps to improve recall and annotation efficiency and performed based on the reference JATS information available from the publisher (as illustrated in Listings~\ref{list:software_direct_xml}, \ref{list:software_manual}, \ref{list:software_website}).
It was performed with a simple implemented tool that displays the reference information and allows a point-and-click annotation.
In the first step it was only annotated if a reference is potentially relevant, and in the second step the citation type was annotated for the marked references. 
Only the citation types \emph{Direct Citation}, \emph{Manual}, and \emph{Website} were considered, as \emph{Software Articles} can generally not be distinguished from other articles based on the reference information alone. 
In the first step, the set of references was reduced to 1,392 references.
The annotation quality of this annotation was measured by assessing the recall based on the known references connected to in-text software mentions that are known to be present in the references.
Overall, 74 of 75 (99\%) known \emph{Direct Citations}, 24 of 25 (96\%) \emph{Manuals}, and 5 of 5 (100\%) \emph{Websites} were successfully identified, which was considered as satisfactory quality. 

\paragraph{Direct Citation Completeness}
was annotated on all available \emph{Direct Citations}, \emph{Manuals}, \emph{Websites}, and \emph{Other} references from the first two annotation step (detailed results for the first annotations are provided in Sections~\ref{sec:res_cit_typ} and~\ref{sec:res_without}; Figure~\ref{fig:pipeline}, middle-right).
The corresponding reference texts were manually extracted from the published PDF documents in February, 2023. 
All information introduced in Section~\ref{sec:ana_cit_comp} was then annotated using the annotation tool BRAT~\citep{stenetorp-etal-2012-brat}.
An example of the annotation is provided in Listing~\ref{list:software_direct_txt}.
The annotation was performed by two annotators and the IAA was initially estimated on 10\% overlap at a value of $\kappa = .82$, which is considered in the range of almost perfect agreement~\citep{landis1977measurement}. 
The remaining data was then annotated by a single annotator, while challenging cases were further discussed throughout the process. 

\paragraph{Database Accuracy} 
was annotated for the same references as the completeness analyses, based on the JATS information available from publishers and on the reference entries provided by Semantic Scholar and Crossref as described in Section~\ref{sec:ana_data_rep} (Figure~\ref{fig:pipeline}, right).
The publishers references were automatically gathered while the corresponding Semantic Scholar and Crossref entries were manually gathered in August, 2022. 
An automatic collection was not possible due to partially missing entries in both databases that hindered precise matching. 
Additionally, multiple entries per reference are in some cases present in Semantic Scholar, which were all extracted and annotated separately.
The extracted information was then annotated for the same information considered for citation completeness, described above.
To assess the quality of data representation and capture potential errors in reference entries specific tags were introduced in the annotation:
\begin{itemize}
    \item \emph{unstructured} marks information that is not labelled within the database, but instead part of a single field containing multiple information about the software. Entries can be partially structured, e.g., the creator and publication date being labelled, but version and software name being unstructured within one field.
    \item \emph{wrong place} indicates that information is structured but with a false underlying concept, e.g., a creator being labelled as a publication venue;
    \item \emph{wrong content} indicates that the information in a database is wrong;
    \item \emph{incomplete content} indicates that some information is only partly presented, and part of the original information is lost;  
    \item \emph{duplicate} indicates duplicate information introduced by a database. Note that this entry refers to duplicate information within one reference, not the duplicate entries for one reference within Semantic Scholar mentioned above. 
\end{itemize}
Since this annotation proved to be challenging from the beginning, all references were annotated by two annotators with discussion of all challenging cases to achieve high quality data. 
As before, the tool BRAT was used for this annotation. 
The result for this annotation is illustrated in Listings~\ref{list:software_direct_xml}, \ref{list:software_direct_sem}, and \ref{list:software_direct_cro}, which illustrate the respective JATS, Semantic Scholar, and Crossref representations. 
\begin{listing}[tb]
    \includegraphics[width=\textwidth]{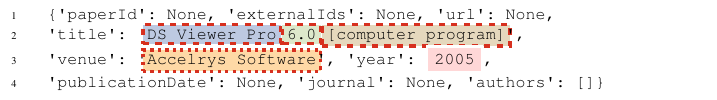}
    \caption{Semantic Scholar reference entry corresponding to the JATS entry in Listing~\ref{list:software_direct_xml} [ID: PMC2134966, Semantic Scholar ID: 1116831, \cite{del2007modular}]. Meta-data is highlighted for: \authorbox{Developer}, \yearbox{Year}, \namebox{Name}, \verbox{Version}, \typebox{Type of Citation}. Meta-data represented in an \unstructuredbox[inner sep=2pt]{unstructured} manner, for which no label is provided, and meta-data represented with the \misrepbox[inner sep=2pt]{wrong label} are marked. Information on the  \urlbox{URL} and \doabox{Date of Access} from the original publisher information is missing.}
    \label{list:software_direct_sem} 
\end{listing}
\begin{listing}[tb]
    \includegraphics[width=\textwidth]{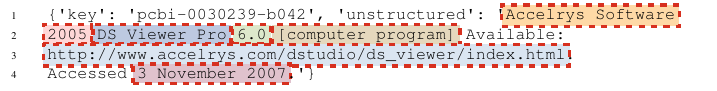}
    \caption{Crossref reference entry corresponding to the JATS entry in Listing~\ref{list:software_direct_xml} [ID: PMC2134966, \cite{del2007modular}]. Meta-data is highlighted for: \authorbox{Developer}, \yearbox{Year}, \namebox{Name}, \verbox{Version}, \typebox{Type of Citation},  \urlbox{URL}, and \doabox{Date of Access}. Meta-data represented in an \unstructuredbox[inner sep=2pt]{unstructured} manner is marked.}
    \label{list:software_direct_cro} 
\end{listing}

\paragraph{Annotation Effort:}
Overall, considerable annotation effort was necessary to generate the high quality dataset described above. 
The annotation of 603 references for citation type took an estimate of 30s per reference, summing to a total of 11h for two annotators and the subsequent re-annotation of 68 references. 
The annotation of $\approx 29,000$ references to identify software citations without plain text mentions of is estimated to take 5s per reference in the first run summing to 40h, and 20s in the second run summing to 4h. 
The annotation of plain text annotation completeness for all \emph{Direct Citations} is estimated with around 2 min per 205 reference, summing to 8h including the overlap for quality control. 
The final annotation of JATS and database entries---including gathering the corresponding entries---is estimated with around 7 min per reference, summing to 48h as all references were examined by two annotators with additional 2h for manual identification of database entries.
In total, this amounts to 113h spent on annotation and quality control. 

\section{Results}
In this section the results for analyses outlined in Section~\ref{sec:analyses} are presented, addressing each of the four main research objectives individually. 
All results are based on the manually annotated data, for which the annotation process is outlined above. 

\subsection{Citation Resource Types}
\label{sec:res_cit_typ}
The resource types for bibliography entries connected to in-text software mentions were systematically annotated to analyze their distribution. 
Overall, 603 entries were annotated based on the original {\sc SoMeSci} annotations of 591 in-text citations connected to software. 
The numbers differs because multiple reference entries can be referenced by one in-text citation string within {\sc SoMeSci}.
For the following analyses, 9 duplicate entries in which the same software was cited twice in one article were excluded. 
Further, all 30 references contained in the {\sc SoMeSci} \emph{Creation Sentences} set were excluded because new research software is developed in their scope and we argue that including them could add a bias as authors publishing software might be more particular about software citation than other authors. 
Moreover, 25 (4\%) additional references had to be excluded from analyses because the bibliography entries were not related to the in-text software mention, even so the citation was directly associated with the software in the article full-text document. 
Further investigation into the underlying reasons showed that 12 (2\%) cases described prior use cases for the software, 7 (1\%) were article errors either by authors or publishers where all citation numbers in the document were mixed up, and 6 (1\%) are entirely unrelated to the software. 

\begin{figure*}[tb]
    \includegraphics[width=\textwidth]{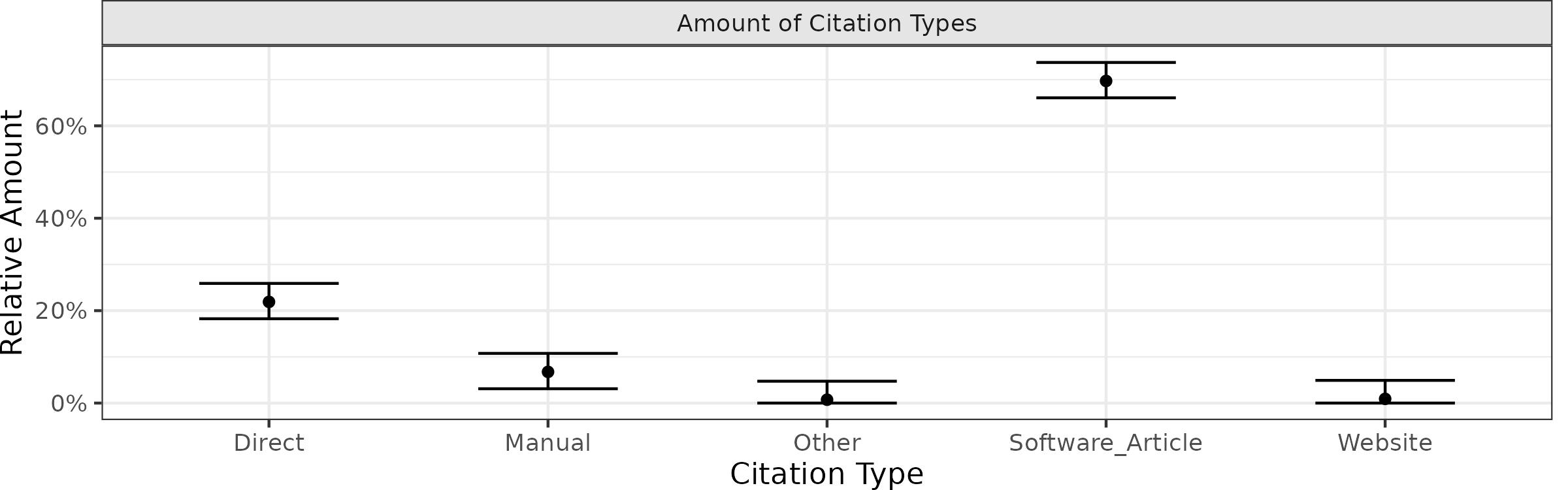}
    \caption{Relative amount of different software citation types with confidence intervals.}
    \label{fig:cite_types}
\end{figure*}

The distribution of resource types as introduced in \ref{sec:ana_formal_citation_type} is illustrated in Figure~\ref{fig:cite_types}. 
The annotation results show that most references are \emph{Software Articles} in 375 or 69.6\% (95\% CI: [65.9, 73.6]) cases, followed by \emph{Direct Citations} in 120 or 22.3\% (95\% CI: [18.6, 26.3]) and \emph{Manuals} in 35 or 6.5\% (95\% CI: [2.8, 10.5]) cases.
\emph{Websites} and \emph{Other} references were only found in 5 or 0.9\% (95\% CI: [0, 5]) and 4 or 0.7\% (95\% CI: [0, 4.8]) cases, respectively. 

\begin{figure*}[tb]
    \includegraphics[width=\textwidth]{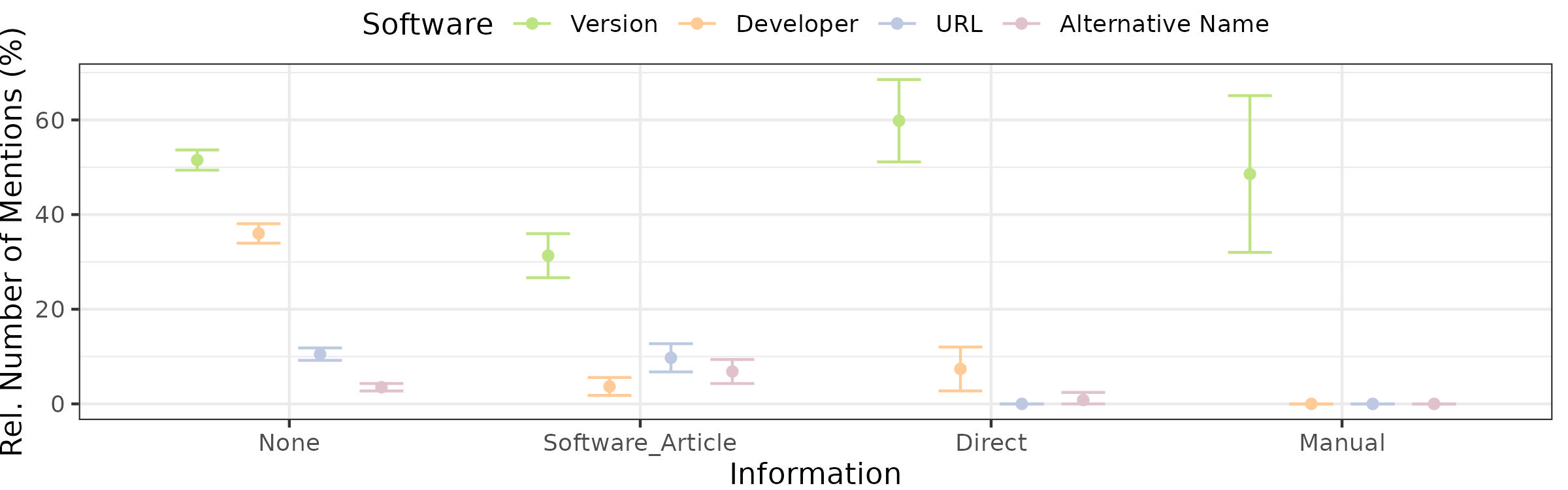}
    \caption{Relative amount of meta-data provided with respect to the formal citation type. Software mentions that were not formally cited are aggregated under label ``None''. 95\% CIs are shown. }
    \label{fig:informal_with_type}
\end{figure*}

It was further investigated if there is an interaction between the citation type and the meta-data provided within the full-text document of a publication because authors might provide the information inherently missing from resource types such as \emph{software article} in the full-text document of a publication. 
Therefore, the amount of stated versions, release dates, developers, and URLs in the full-text document is compared with respect to the type of formal citation, including all mentions that were not formally cited as an additional class.  
Versions and releases are summarized under version if at least one of both is given because releases are rare in the {\sc SoMeSci} dataset.
Further, the types \emph{Website} and \emph{Other} are excluded because there are too few data points for them. 
The results are illustrated in Figure~\ref{fig:informal_with_type}.
The results show that less versions are mentioned with software articles with 31.3\% (95\% CI: [26.7, 36.0]) as compared to mentions without formal citations with 51.5\% (95\% CI: [49.4, 53.7]) but also in comparison with direct citations with 59.8\% (95\% CI: [51.1, 68.5]). 
We employed a chi-square test, $\chi^2(1,N{=}2482){=}51.8,p{<}.001$, to test whether the number of provided versions systematically differs between not cited software and software cited by a software article, and use Cramer's V to estimate the effect size, $V{=}0.15$. 
The test shows that significantly less versions are mentioned in-text when software is cited by a software article with a small effect size. 
Developers are mentioned less in all cases where software is formally cited (direct 7.4\% (95\% CI: [2.7, 12.0]), manual 0\%, software articles 3.7\% (95\% CI: [1.8, 5.6])) as compared to not formally cited software (36\% (95\% CI: [34.0, 38.1]), however, in most citation types developer attribution is given, including software articles. 
The in-text mention of URL is at a similar level between software cited with software articles and software that was not cited, while URLs have never been provided in-text when software was cited directly or through a manual.
A similar picture is found in alternative names. 

\subsection{Formal Software Citation without in-text Software Mentions}
\label{sec:res_without}
All references within the {\sc SoMeSci} \emph{PLoS methods} and \emph{PubMed fulltexts} sets were analyzed to determine if software is formally cited without being mentioned in the article's full-text document. 
All contexts of identified references were further manually examined to determine the reasons why the software was not mentioned in-text. 
In total, 32 formal software citations were identified within the references of all articles. 
However, 11 of these citations are actually connected to in-text software citations but appear in non-annotated parts of the {\sc SoMeSci} \emph{PLoS Methods} set. 
The remaining 21 are not connected to in-text software and are contained within 17 articles. 
The closer examination of the corresponding reference contexts revealed that 2 of the cases are due to annotation errors in the original {\sc SoMeSci} data, and 5 can be attributed to errors within the articles, e.g., mixed up references as described in Section~\ref{sec:res_cit_typ}. 

The remaining 14 cases reflect the citation practice of interest where software is formally cited without in-text mention and consist of 7 \emph{Direct Citations}, 4 \emph{Manuals} and 3 \emph{Websites}.
To assess the extend of this practice, these numbers are considered in relation to the number of overall formal citations within the analyzed articles from \emph{PLoS methods} and \emph{PubMed fulltexts}.
This amounts 8.5\% of \emph{Direct Citations} for a total of 82 \emph{Direct Citation} with 75 cases where the software is mentioned in-text, 13.8\% of \emph{Manuals} with 29 overall cases and 25 in-text mentions, and 37.5\% of \emph{Websites} with 8 overall cases and 5 in-text mentions. 
Note that the sample size for \emph{Manuals} and \emph{Websites} is quite small and that all additionally identified \emph{Websites} result from the same article.
The manual analyses of the underlying citation practices showed that in 4 cases generic terms were mentioned instead of the software name (e.g., ``processing was done with [19]'', where [19] is the software citation), in 7 cases the use of a software was not indicated at all, in 3 cases knowledge from the software was cited, e.g., the FAQ of the software being referenced instead of the software. 

\subsection{Citation Completeness}
\label{sec:res_cit_comp}
\begin{figure*}[tb]
    \includegraphics[width=\textwidth]{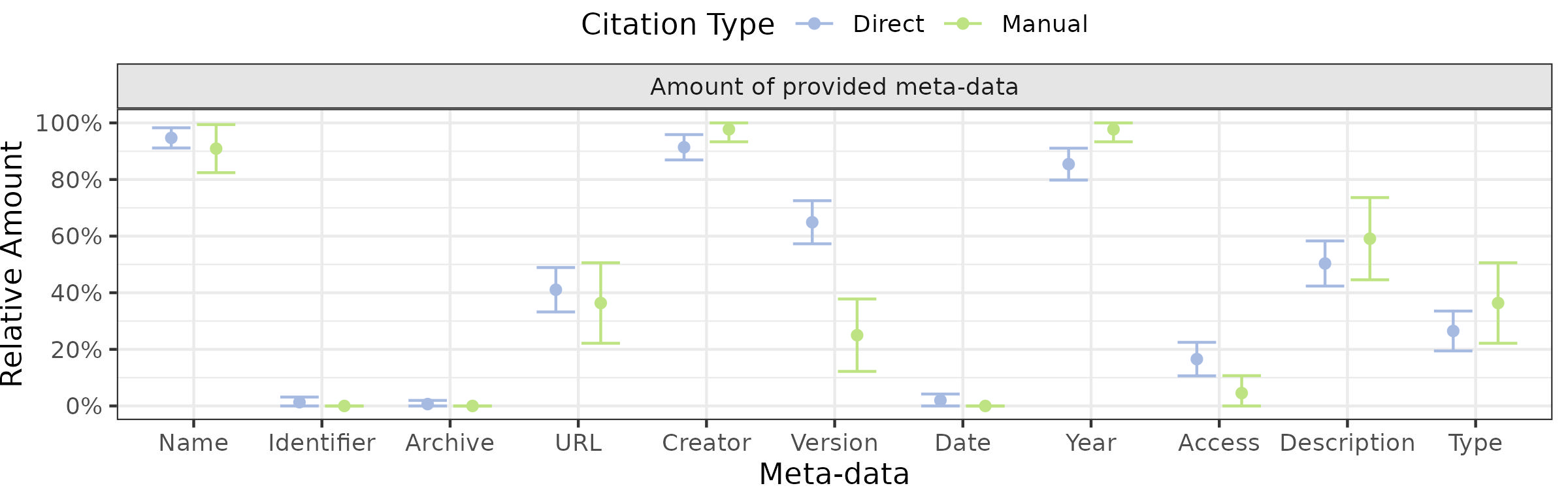}
    \caption{Relative citation completeness of software meta-data in \emph{Direct Software Citations} and \emph{Software Manuals} with 95\% CIs.}
    \label{fig:cite_comp}
\end{figure*}

The completeness of formal software citations was analyzed for 153 \emph{Direct Citations}---including 8 \emph{Website} and 4 \emph{Other} citations as incomplete \emph{Direct Citations}---and 44 \emph{Manuals} identified in Section~\ref{sec:res_cit_typ} and~\ref{sec:res_without}, but as in Section~\ref{sec:res_cit_typ} excluding 8 references from the \emph{Creation sentences} set.
The corresponding results are summarized in Figure~\ref{fig:cite_comp}.
Regarding \emph{Direct Citations}, we found that the \emph{Name}, \emph{Creator} and \emph{Publication Year} of software are commonly mentioned in 146 or 94.8\% (95\% CI: [91.3, 98.3]), 141 or 91.6\% (95\% CI: [87.2, 95.9], and 132 or 85.7\% (95\% CI: [80.2, 91.2]) instances, respectively. 
\emph{Version}, \emph{Description}, and \emph{URL} are less common with 100 or 64.9\% (95\% CI: [57.4, 72.5]), 77 or 50\% (95\% CI: [42.1, 57.9]), and 62 or 40.3\% (95\% CI: [32.5, 48.0]) of instances, while the \emph{Type of Citation} and \emph{Date of Access} are only rarely provided in 44 or 28.6\% (95\% CI: [21.4, 35.7]) and 25 or 16.2\% (95\% CI: [10.4, 22.1]) of cases. 
\emph{Release date} (3, 2\% with 95\% CI: [0, 4.13]), \emph{Identifier} (1, 0.6\% with 95\% CI:[0, 1.92]), and \emph{Archive} (1, 0.6\% with 95\% CI:[0, 1.92]) were only sporadically found.  
Regarding \emph{Manuals}, most results are at a comparable level, with a difference in \emph{Version}, which are less often contained in \emph{Manual} citations with 25\% (95\% CI: [12.2, 37.8]). 
Exact results for \emph{Manuals} are provided in the Supplements\footnote{\label{foot:supplements}Available at \url{https://github.com/dave-s477/SoMeSci_Citation}}. 

\begin{figure*}[tb]
    \includegraphics[width=\textwidth]{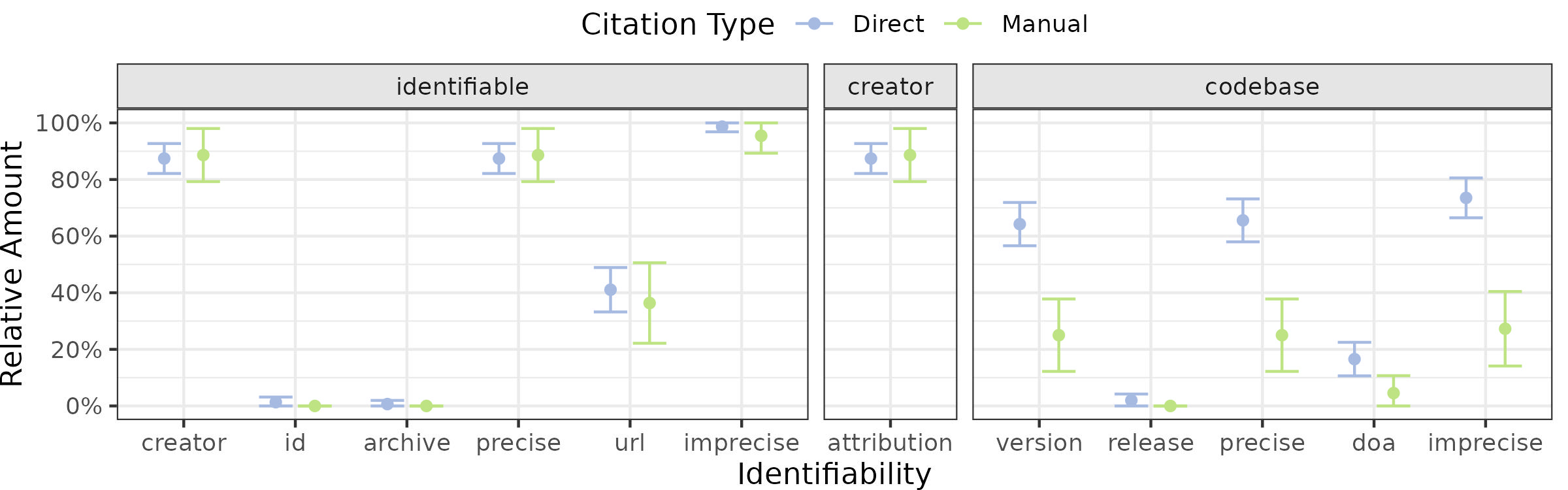}
    \caption{Relative citation completeness of \emph{Direct Software Citations} and \emph{Software Manuals} with respect to software identification, author attribution, and code base identification, including 95\% confidence intervals. Regarding software identification, ``precise'' describes cases in which software is clearly identifiable, while ``imprecise'' includes cases where only a \emph{URL} is provided as identifier, which is considered as unsuited to clearly identify a software with details outlined in Section~\ref{sec:ana_cit_comp}. Regarding codebase identification, the categories are defined similarly, with ``imprecise'' including cases where only a \emph{Date of Access} is provided.}
    \label{fig:comp_id}
\end{figure*}

The analysis was extended to further cover whether software is identifiable, if the creator can be attributed, and whether the code base can be identified, as defined in Section~\ref{sec:ana_cit_comp}.
The corresponding results are illustrated in Figure~\ref{fig:comp_id}.
Regarding \emph{Direct Citation}, Software can be identified with high confidence in 132 or 87.4\% (95\% CI: [82.1, 92.7]) of cases based on the mention of \emph{Name} and \emph{Creator}.
\emph{Archive} and \emph{Identifier} are only stated in one case each and overlap with mentions of \emph{Name} and \emph{Creator}.
Furthermore, if cases where a \emph{URL} is provided are considered as identifiable the overall number of identifiable cases increases to 149 or 98.7\% (95\% CI: [96.9, 100]). 
However, URLs are often not persistent or might only point the developer instead of the software, which makes it dangerous to assume that they are always identifiable. 
The creator can be attributed in 132 or 87.7\% (95\% CI: [82.5, 92.9]) of cases, in which he was provided. 
The exact code bases can be identified with high confidence in 101 or 65.6\% (95\% CI: [58.1, 73.1]) of cases, with \emph{Versions} being provided in 99 cases and \emph{Release Dates} in three with one case overlap. 
Note that the software itself also has to be identifiable to identify the code base.
Therefore, the numbers for code base identification are based on the 98.7\% of software that was found to be identifiable before.
Considering a \emph{Date of Access}, available in 25 cases, sufficient for code based identification under the assumptions that the newest available version at the date of access was used, the code base can be identified in 113 or 73.4\% (95\% CI: [66.4, 80.4]) of cases. 
Regarding \emph{Manual} citations, the values for software identification and creator attribution are at equal levels, however, the value for code base identification is lower with only 25\% (95\% CI: [12.2, 37.8]) and 27.3\% (95\% CI: [14.1, 40.4]) being identifiable with and without considering a \emph{Date of Access}, respectively. 
This value is mostly caused by the lower number of provided versions as outlined above.
Exact results for \emph{Manuals} are provided in the Supplements\footref{foot:supplements}. 

Same as in Section~\ref{sec:res_cit_typ}, the completeness was further investigated including meta-data provided with in-text software mention.
It is possible that meta-data describing a software is provided in the full-text document instead of the software citation. 
While this would have the drawback of making the in-text information not directly identifiable, it would still mean that the required information to describe the software has been provided within an article.
Therefore, the formally and informally provided information is compared and aggregated to determine if completeness can be gained by observing both.
This was only performed for samples that have an in-text mention and are annotated in {\sc SoMeSci}, therefore, excluding the samples described in Section~\ref{sec:res_without}.
Further, only the meta-data of \emph{Version/Release}, \emph{Creator}, and \emph{URL} is considered as the remaining information does not overlap between the annotations. 
The results are given in supplementary Figure S5.
By definition, the software \emph{Name} is always given for informal mentions in the {\sc SoMeSci} annotation, therefore, the number of overall provided names within theses samples is 100\%. 
\emph{Creators} are only rarely provided when software is cited with a \emph{Direct Citation} in 6.9\% (95\% CI: [2.6, 11.3]) of cases and for \emph{Manuals} in 0 cases. 
All of the cases where a developer was mentioned in-text overlapped with cases where the \emph{Creator} was also provided in the formal references, therefore, not improving the overall coverage. 
\emph{Versions} are provided quite often with the informal mention when software is cited with a \emph{Direct Citation} with 58.5\% (95\% CI: [50, 66.9]) and 47.1\% (95\% CI: [30.3, 63.8]) for \emph{Manual} citations. 
Further, through aggregating over formal and informal information the overall coverage for versions improves up to 80\% (95\% CI: [73.1, 86.9]) for \emph{Direct Citations} and to 55.9\% (95\% CI: [39.2, 72.6]) for \emph{Manuals}. 
\emph{URLs} were never mentioned in the full-text document when software was formally cited by either a \emph{Direct Citation} or a \emph{Manual}. 

\subsection{Database Accuracy}

The quality of database representation was evaluated on the same references as the completeness and the set of references in the \emph{Creation Sentences}, because only the representation of the information is investigated here not its amount.
As outlined in Section~\ref{sec:anno}, it is investigated which information is available from the different databases, whether all available information is covered, how it is structured, and whether it is correct. 
The quality of database representation was investigated individually for the considered information, e.g., \emph{Name, Version, Developer}, and illustrated in newly established, adapted alluvial plots that illustrate and compare the availability, structure, and correctness of individual references between the publisher's JATS information, Semantic Scholar, and Crossref. 
Potential errors that can occur in a database representation are illustrated in Listing~\ref{list:database_error}, including unstructured representation, incomplete representation of information, errors in information, and addition of wrong information.
\begin{listing}[tb]
    \includegraphics[width=\textwidth]{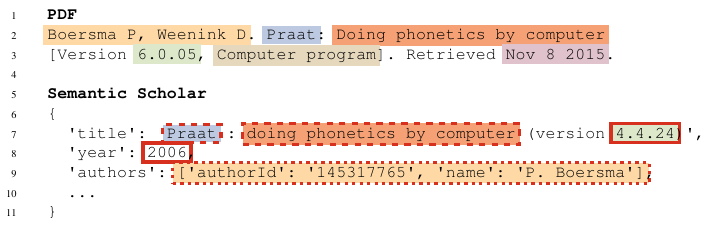}
    \caption{Original PDF reference information and the corresponding Semantic Scholar reference entry for \emph{Direct Software Citation} in the presented dataset [ID: PMC5574543, Semantic Scholar ID: 27589464, ~\cite{wood2017towards}]. Several errors are present in Semantic Scholar: information on date of access and citation type are lost, the content for version is wrong, author information was partially lost and wrong information on the publication year is added. Meta-data is highlighted for: \authorbox{Developer}, \namebox{Name}, \descbox{Description}, \verbox{Version}, \typebox{Type of Citation}, and \doabox{Date of Access}. Meta-data represented in an \unstructuredbox[inner sep=2pt]{unstructured} manner, with \incompletebox[inner sep=1pt]{incomplete} information, or \wrongbox[inner sep=2pt]{errors} is marked. Note that the version in Semantic Scholar is both unstructured and wrong.}
    \label{list:database_error}  
\end{listing}
As described in Section~\ref{sec:ana_data_rep} multiple entries for one reference can exist in Semantic Scholar.
Overall, 40 (24.8\%) out of 161 represented citations have duplicate entries, with 33 (27\% of 121) in \emph{Direct Citations} and 7 (17.5\% of 40) in \emph{Manuals}. 
For all following analyses the most complete entry, covering most relevant information, was selected. 

\subsubsection{Database Errors}

Before analyzing the individual information some general analyses were performed. 
In both databases missing entries for references were identified, where it is necessary to distinguish two cases of missing references: entire articles missing and individual references missing. 
References that are missing because the entire articles is not contained in a database are ignored because this is a problem of overall coverage and does not provide any information about the quality of software citation representation. 
However, individual references missing, even so an article is represented, can point to a problem regarding software citation representation and needs to be investigated.
Overall, 36 of 157, 22.9\% (95\% CI: [16.4, 29.5]) of \emph{Direct Citations} and 5 of 45, 11.1\% (95\% CI: [1.93, 20.3]) of \emph{Manual} references were individually missing from Semantic Scholar. 
In Crossref 8 of 155, 5.2\% (95\% CI: [1.7, 8.6]) of \emph{Direct Citations} and 6 of 47, 12.8\% (95\% CI: [3.2, 22.3]) of \emph{Manuals} were missing\footnote{Note that the overall number of references differs between databases because they differ in the number of references ignored because entire articles are missing.}. 
To investigate if this is a systematic bias concerning software citations, we further investigated what amount of \emph{Software Articles} are missing from the databases, serving as a sample of regularly published articles.
We identified 4 of 382, 1.1\% (95\% CI: [0, 2.1]) and 1 of 381, 0.3\% (95\% CI: [0, 1]) of \emph{Software Articles} missing from Semantic Scholar and Crossref, respectively.
To test if the amount of missing articles differs between \emph{Software Articles} and \emph{Direct Software Citation} we employed a chi-squared test for each of the databases, Semantic Scholar $\chi^2(1,N{=}539){=}74.4,p{<}.001$ and Crossref $\chi^2(1,N{=}536){=}13.2,p{<}.001$ with effect sizes of $V{=}0.38$ for Semantic Scholar and $V{=}0.17$ for Crossref, estimated by Cramer's V.
We do not employ further tests regarding software manuals because there is fewer data available and statements would be less reliable. 

Furthermore, it was observed during annotation that Semantic Scholar sometimes adds wrong information without relation to the original reference information (see Listing~\ref{list:database_error}), and that correct information is in some cases duplicated in a wrong location, e.g., the software name is represented as both title and publication venue. 
In total, wrong information is added to reference representations of \emph{Direct Citations} in 19, 15.7\% (95\% CI: [9.2, 22.2]) and for \emph{Manuals} in 25, 62.5\% (95\% CI: [47.55, 77.5]), while duplicate information is added in 26, 21.5\% of \emph{Direct Citations} and 2, 5\% of \emph{Manuals}, with 3 cases overlapping.
Both problems were not observed for Crossref.

\subsubsection{Presentation of Results}
\begin{figure*}[tb]
    \includegraphics[width=\textwidth]{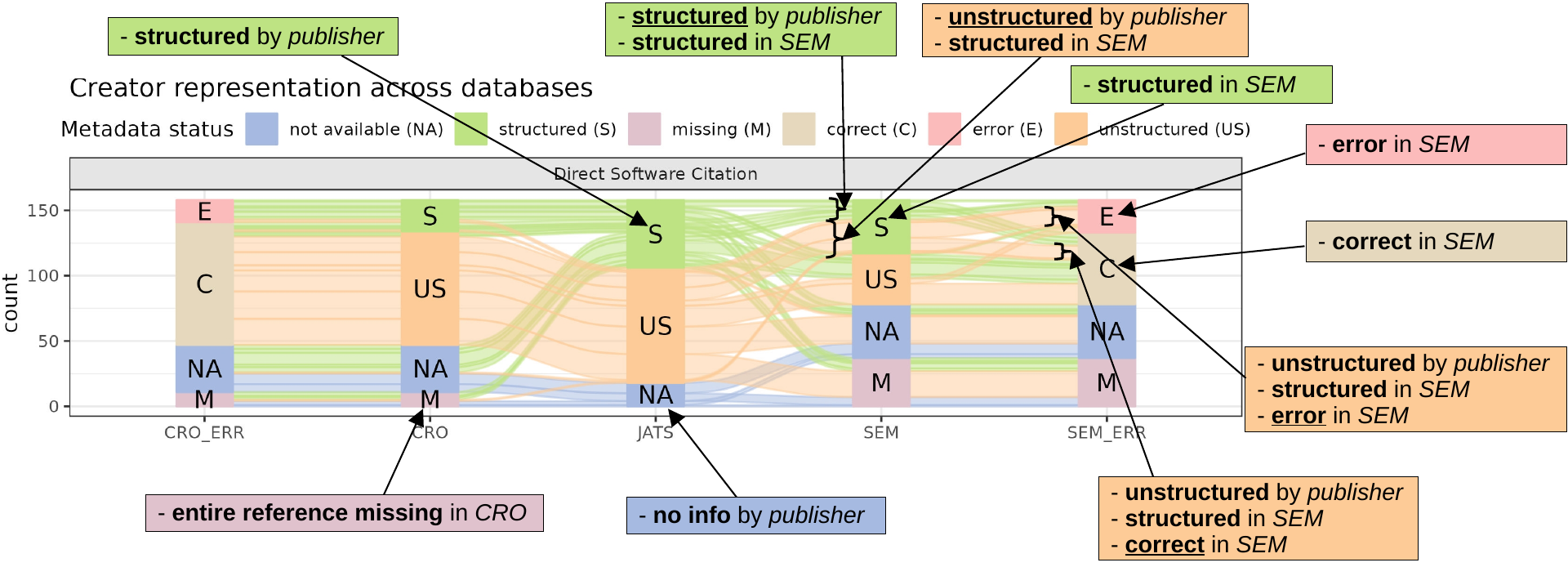}
    \caption{Adapted alluvial plot with added explanation of information flow. The detailed plot design is described in the text.}
    \label{fig:plot_ex}
\end{figure*}

In the following the individual information is illustrated through adapted alluvial plots, that are introduced in the following.
An example plot illustrating the adapted alluvial plot is given in Figure~\ref{fig:plot_ex}.
All annotated samples are individually listed in the plot from top to bottom while their order can change from left to right. 
The flow of a specific sample is indicated by the color originating from the middle column named \emph{JATS}.
If multiple samples have the same information flow their lines are summarized. 
The middle \emph{JATS} columns shows the information available from the publisher, and indicates whether the information is available and whether it is correctly structured. 
The columns left and right from the middle show the same information for Crossref (\emph{CRO}) and Semantic Scholar (\emph{SEM}), respectively.
The outermost columns \emph{CRO\_ERR} and \emph{SEM\_ERR} show whether the represented information is correct or whether an error is present, for Crossref and Semantic Scholar, respectively.
Since we observed that some references are entirely missing in Crossref and Semantic Scholar they are shown with the special label ``missing'' to indicate that no information is available in these cases. 
This information flow illustration allows to directly compare how the different sources structure the meta-data and whether errors are introduced. 
Particularly, the difference between the structure provided by the publisher and the corresponding representation by the databases can be observed. 

\subsubsection{Software Name}
The results for database accuracy of software names are given in Figure~\ref{fig:db_name}, and a further summary of the results is provided in Appendix Table~\ref{tab:res_summary}.
\begin{figure*}[tb]
    \includegraphics[width=\textwidth]{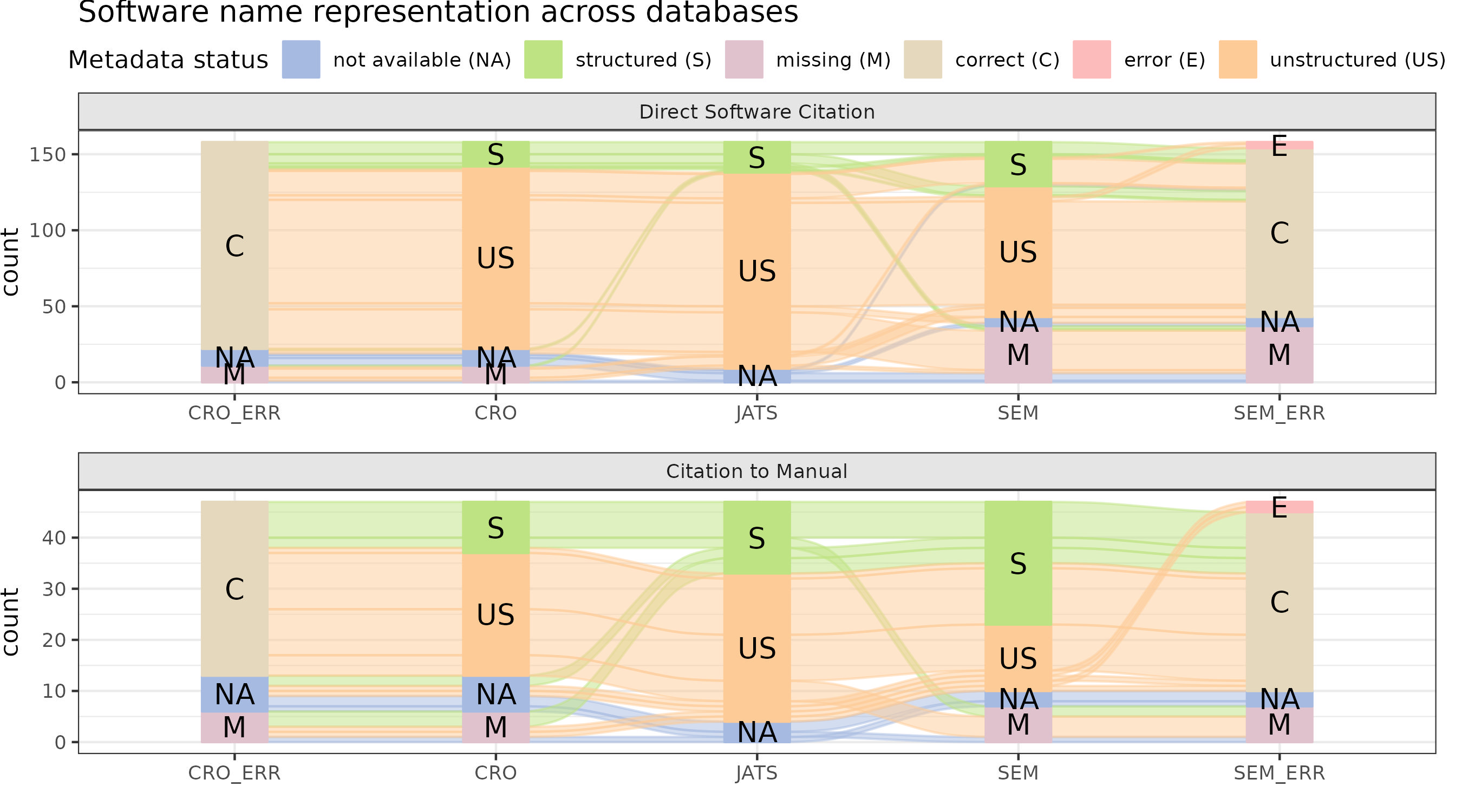}
    \caption{Adapted alluvial plot illustrating the availability, structure, and correctness of software name representations within \emph{Direct Software Citations}. The results for availability and structure for the publisher's JATS in the middle column are compared to the results for Crossref (CRO) on the left, Semantic Scholar (SEM) on the right. The outermost columns show the correctness for Crossref (CRO\_ERR) on the left and Semantic Scholar on the right (SEM\_ERR). Missing samples in Crossref and Semantic Scholar are added in all columns for completeness. The illustration follows the principle illustrated in Figure~\ref{fig:plot_ex}.}
    \label{fig:db_name}
\end{figure*}
The software name is commonly included by publishers in both \emph{Direct Citation} (94.3\%) and \emph{Manual} citations (91.5\%), with only some information represented in a structured manner with 12.7\% of \emph{Direct Citations} and 29.8\% of \emph{Manuals}. 
Crossref and Semantic Scholar only loose information on software names in rare cases with 2\% and 3.3\%, respectively, for \emph{Direct Citations} and 9.8\% and 0\% for \emph{Manuals}. 
In turn, information is added by Semantic Scholar in 0.8\% of \emph{Direct Citations}. 
Semantic Scholar manages to increase the ratio of structured information, for both \emph{Direct Citations} (24\%) and \emph{Manuals} (60\%), with structured samples outweighing unstructured samples for \emph{Manuals}, while Crossref directly reflects publisher structure, when information is not lost\footnote{The results are always reported excluding entirely missing references (M).}. 
Notably, Semantic Scholar does not retain structure for all \emph{Direct} references, but instead looses structure for 5.8\%, and adds structure for 14.9\% of references. 
Regarding \emph{Manuals}, Semantic Scholar does not loose structure, but adds it in 30\% of cases.
All information on software names contained in Crossref is correct, while Semantic Scholar introduces a small amount of errors in both \emph{Direct Citations} (3.5\%) and \emph{Manuals} (5.4\%\footnote{The amount of errors is always reported excluding not covered information (NA).}. 
Regarding \emph{Manuals}, all errors are due to misrepresentation of software names as other information, while for \emph{Direct Citations} 75\% of errors are due to misrepresentation and 25\% due to wrong information.

\subsubsection{Creator}
The results for database accuracy of software creators are given in Figure~\ref{fig:db_creator}, and a further summary of the results is provided in Appendix Table~\ref{tab:res_summary}.
\begin{figure*}[tb]
    \includegraphics[width=\textwidth]{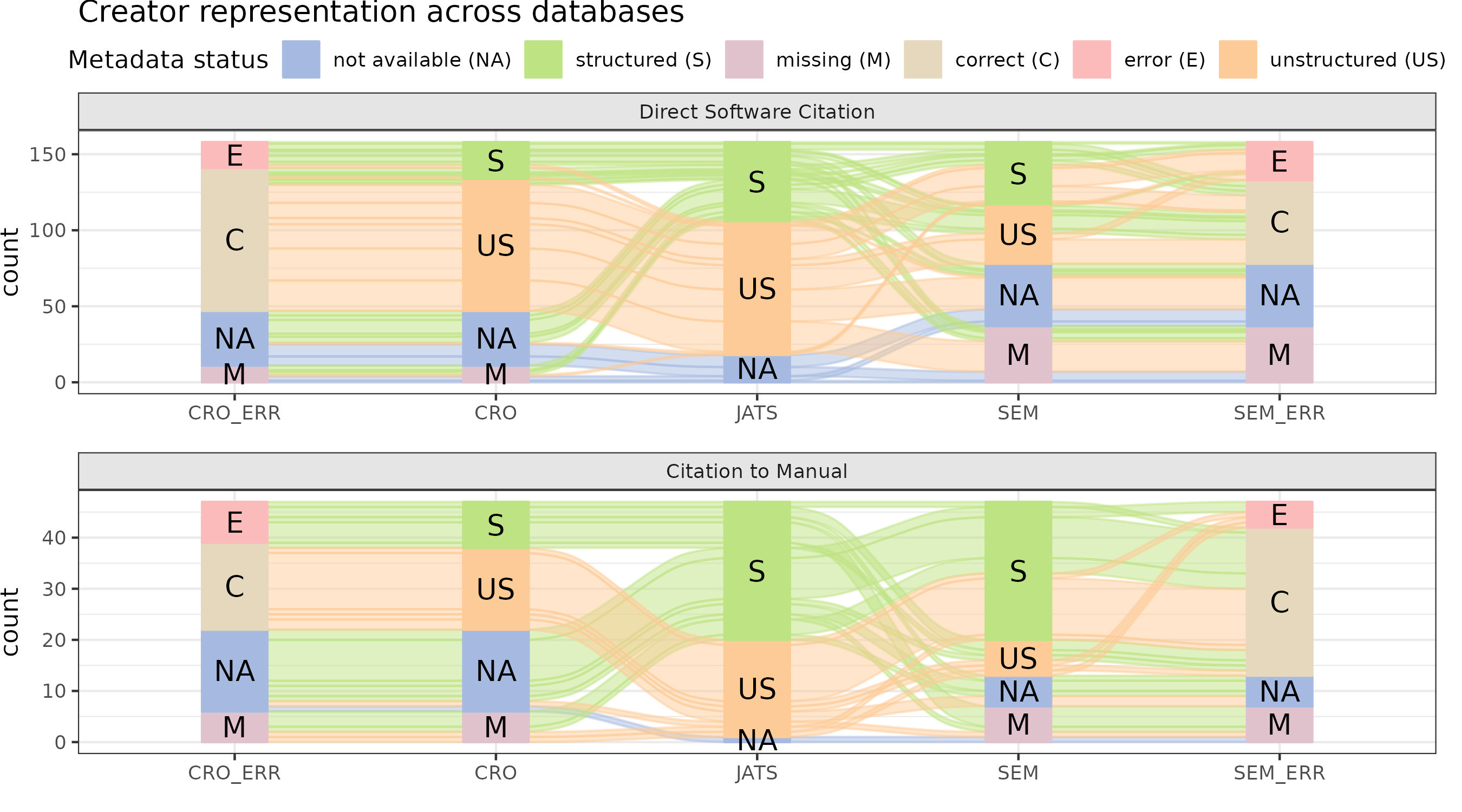}
    \caption{Alluvial plot illustrating the availability, structure, and correctness of software creators representations within \emph{Direct Software Citations} and \emph{Manuals} following the principle of Figure~\ref{fig:plot_ex}.}
    \label{fig:db_creator}
\end{figure*}
The software creator is commonly included by publishers in both \emph{Direct Citations} (88.6\%) and \emph{Manual} citations (97.9\%). 
It is structured in a majority of \emph{Manuals} (57.4\%) but less often in \emph{Direct Citations} (32.9\%). 
Crossref and Semantic Scholar both loose information on software creator in a notable amount of cases for \emph{Direct Citations} (15\% and 24.8\%, respectively) and \emph{Manuals} (36.6\% and 15\%).
Semantic Scholar manages to increase the ratio of structured information slightly for \emph{Direct Citations} to a value of 33.9\% and strongly for \emph{Manuals} up to 67.5\%, with structured samples clearly outweighing unstructured samples for \emph{Manuals}.
Same as for software name, Crossref mostly reflects publisher structure for creators, when information is not lost. 
Again, Semantic Scholar does not retain structure for all \emph{Direct Citation} references, but instead looses structure for 15.7\%, and adds structure for 21.5\% of references. 
Regarding \emph{Manuals}, Semantic Scholar also looses structure in 10\% but adds it in 32.5\% of references.
Semantic Scholar introduces a notable amount of errors in both \emph{Direct Citations} (31.2\%) and \emph{Manuals} (14.7\%). 
For \emph{Direct Citations} they are distributed between wrong information (44\%), incomplete entries (32\%), and misrepresentation (28\%)\footnote{Errors are not exclusive, therefore, the percentages do not need to sum to 1.}, and for \emph{Manuals} between misrepresentation (80\%) and incomplete entries (20\%).
Crossref also introduces a notable amount of errors in both \emph{Direct Citations} (15.3\%) and \emph{Manual} (32\%) references.
In Crossref almost all errors (\emph{Direct Citation} 94.1\%, \emph{Manual} 100\%) are due to incomplete entries because Crossref only includes the first author when representing article references. 
For articles covered in Crossref the full author information can then be gathered from the article entry corresponding to the reference, but for \emph{Direct Software Citations} and \emph{Software Manuals} this can result in a loss of information due to missing persistent identifiers. 

\subsubsection{Identifier}
Information on the publication venue of a software is analyzed as a combination of ID, Archive Link, and URL, where the most relevant information is chosen in the given order if available\footnote{All meta-data is semantically related and samples for ID and Archive are too rare to analyze individually.}. 
The results for database accuracy of software identifiers are given in Figure~\ref{fig:db_url}, and a further summary of the results is provided in Appendix Table~\ref{tab:res_summary}.
\begin{figure*}[tb]
    \includegraphics[width=\textwidth]{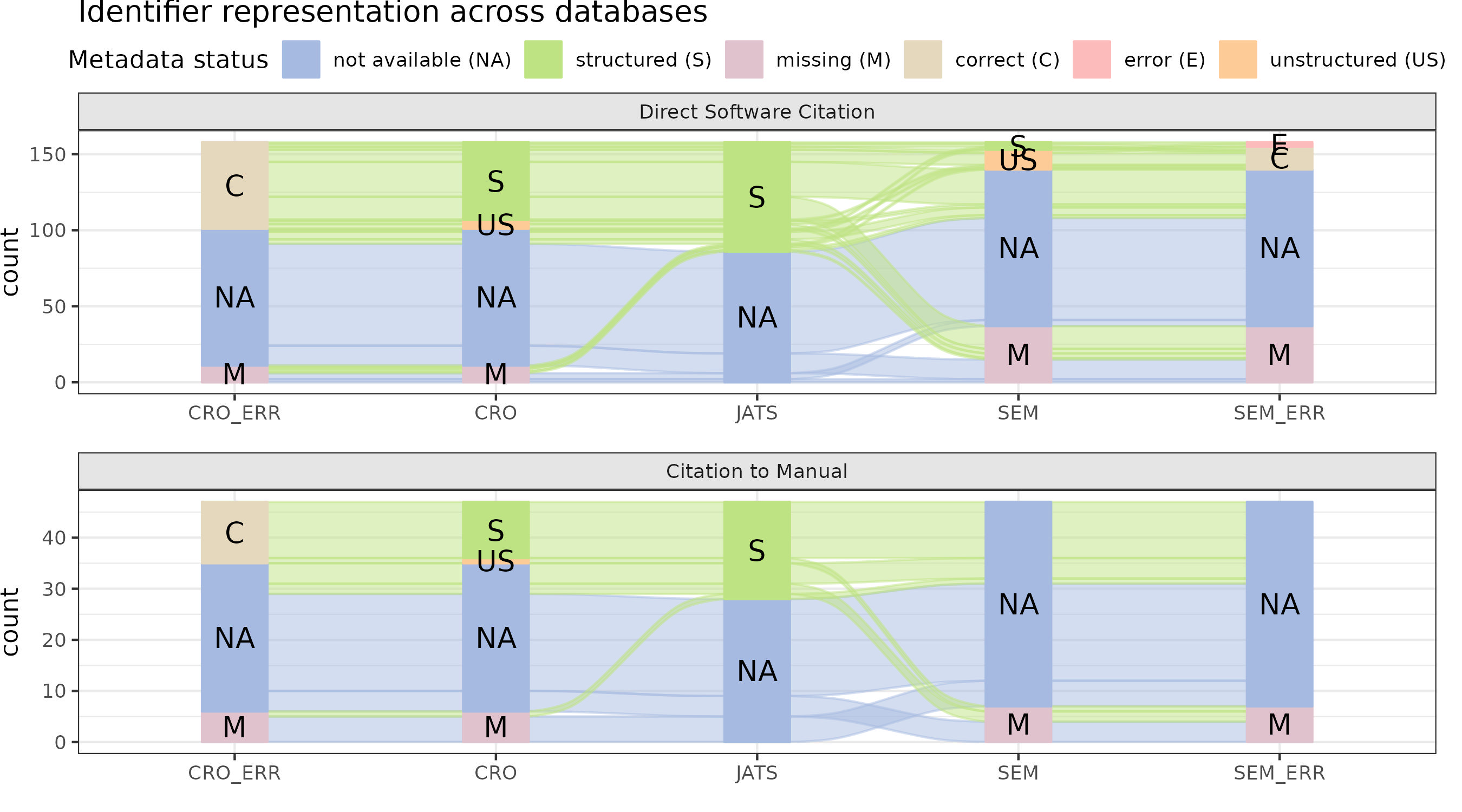}
    \caption{Alluvial plot illustrating the availability, structure, and correctness of software identifier representations within \emph{Direct Software Citations} and \emph{Manuals} following the principle of Figure~\ref{fig:plot_ex}.}
    \label{fig:db_url}
\end{figure*}
A software identifier is included by the publisher in almost half of references in both \emph{Direct Citation} (45.6\%) and \emph{Manual} citations (40.4\%), always in a structured manner.
Semantic Scholar looses information of identifier in a high amount of cases for \emph{Direct Citations} (26.4\%) and always looses it for \emph{Manuals}.
Crossref looses information in fewer cases with 6.8\% for \emph{Direct} and 14.6\% for \emph{Manual}. 
Further, Semantic Scholar looses structure for 10.7\% of \emph{Direct Citations}, while Crossref looses structure for 4.1\% of \emph{Direct Citations} and 2.4\% of \emph{Manuals}.
Errors are only present in rare cases concerning Semantic Scholar and \emph{Direct Citations}, affecting 16.7\% of covered references. 
The errors are due to misrepresentation in 33.3\% of cases and wrong information in 66.7\%.

\subsubsection{Version}
The results for database accuracy of software versions are given in Figure~\ref{fig:db_version}, and a further summary of the results is provided in Appendix Table~\ref{tab:res_summary}.
\begin{figure*}[tb]
    \includegraphics[width=\textwidth]{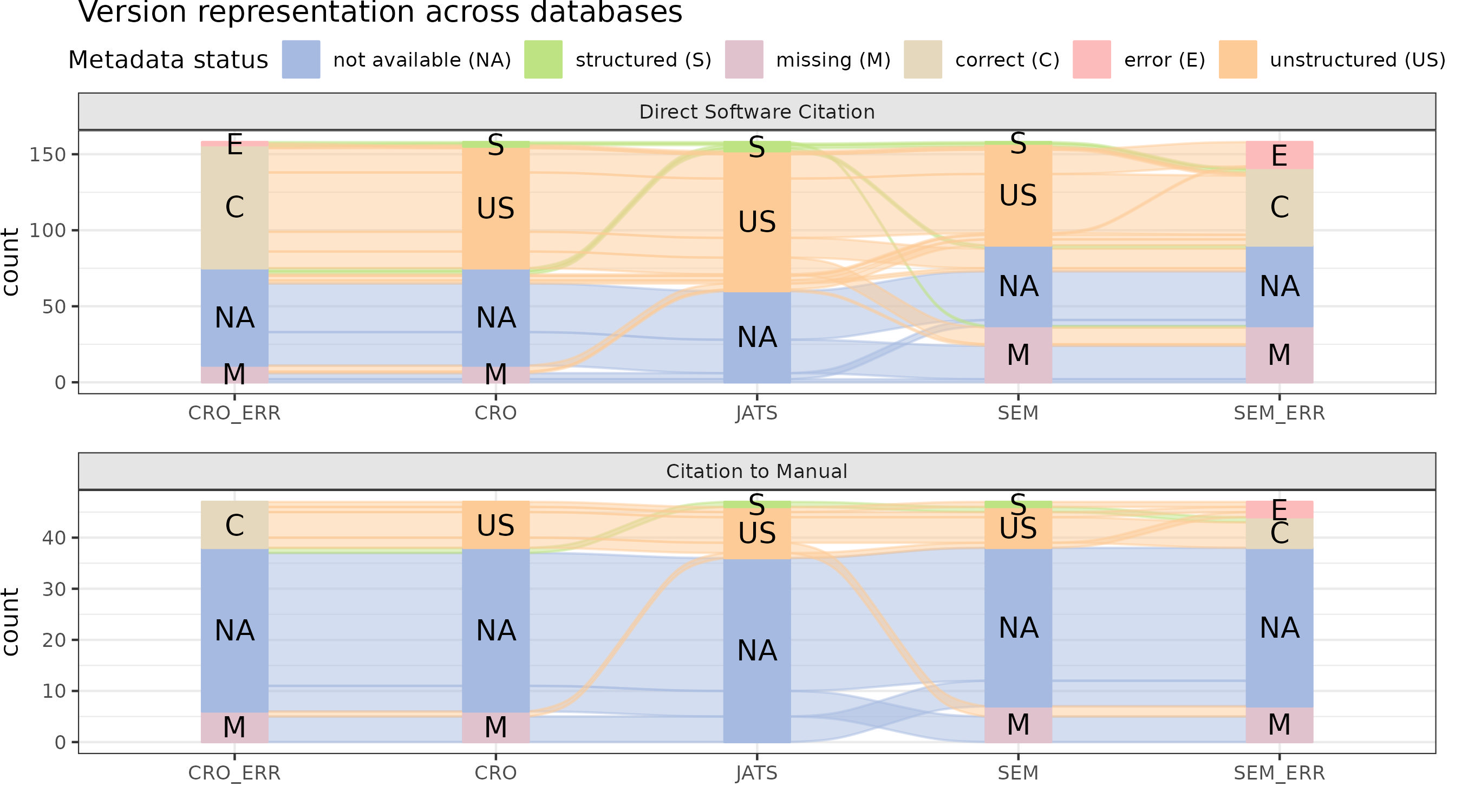}
    \caption{Alluvial plot illustrating the availability, structure, and correctness of software version representations within \emph{Direct Software Citations} and \emph{Manuals} following the principle of Figure~\ref{fig:plot_ex}.}
    \label{fig:db_version}
\end{figure*}
The software version is commonly included by publishers in \emph{Direct Citations} (62\%), but less frequently in \emph{Manual} citations (23.4\%).
For both citation types, versions are rarely represented in a structured manner with 3.8\% in \emph{Direct Citations} and 2.1\% in \emph{Manuals}. 
Crossref rarely looses information on software names in 6.8\% of \emph{Direct} and 2.4\% of \emph{Manuals}. 
Semantic Scholar, on the other hand, looses version information in a considerable amount of \emph{Direct Citations} (14\%), but never in \emph{Manuals}.
Crossref does not loose structure information when samples are represented, but adds structure for 0.7\% \emph{Direct Citations}.
Semantic Scholar looses structure for 1.7\% \emph{Direct Citations} and 2.5\% \emph{Manuals}. 
No errors are present in Crossref for \emph{Manuals} and only few for \emph{Direct Citations} (2.4\%), all due to misrepresentation of the version as other information. 
For Semantic Scholar a notable amount of errors is present in \emph{Direct Citations} (25\%) and \emph{Manuals} (33.3\%). 
The errors in Semantic Scholar for \emph{Direct Citation} are mainly due to wrong information (76.5\%), followed by incomplete information (17.6\%), and misrepresentation (5.9\%), while for \emph{Manuals} they are due to incomplete information (100\%) and misrepresentation (66.7\%).

\subsubsection{Release Dates}
Same as for the identifier, information on release dates is summarized for analyses, where the release date, date of access, and publication year are prioritized for analyses in the given order to always select the most complete information. 
The results for database accuracy of release date information are given in Figure~\ref{fig:db_year}, and a further summary of the results is provided in Appendix Table~\ref{tab:res_summary}.
\begin{figure*}[tb]
    \includegraphics[width=\textwidth]{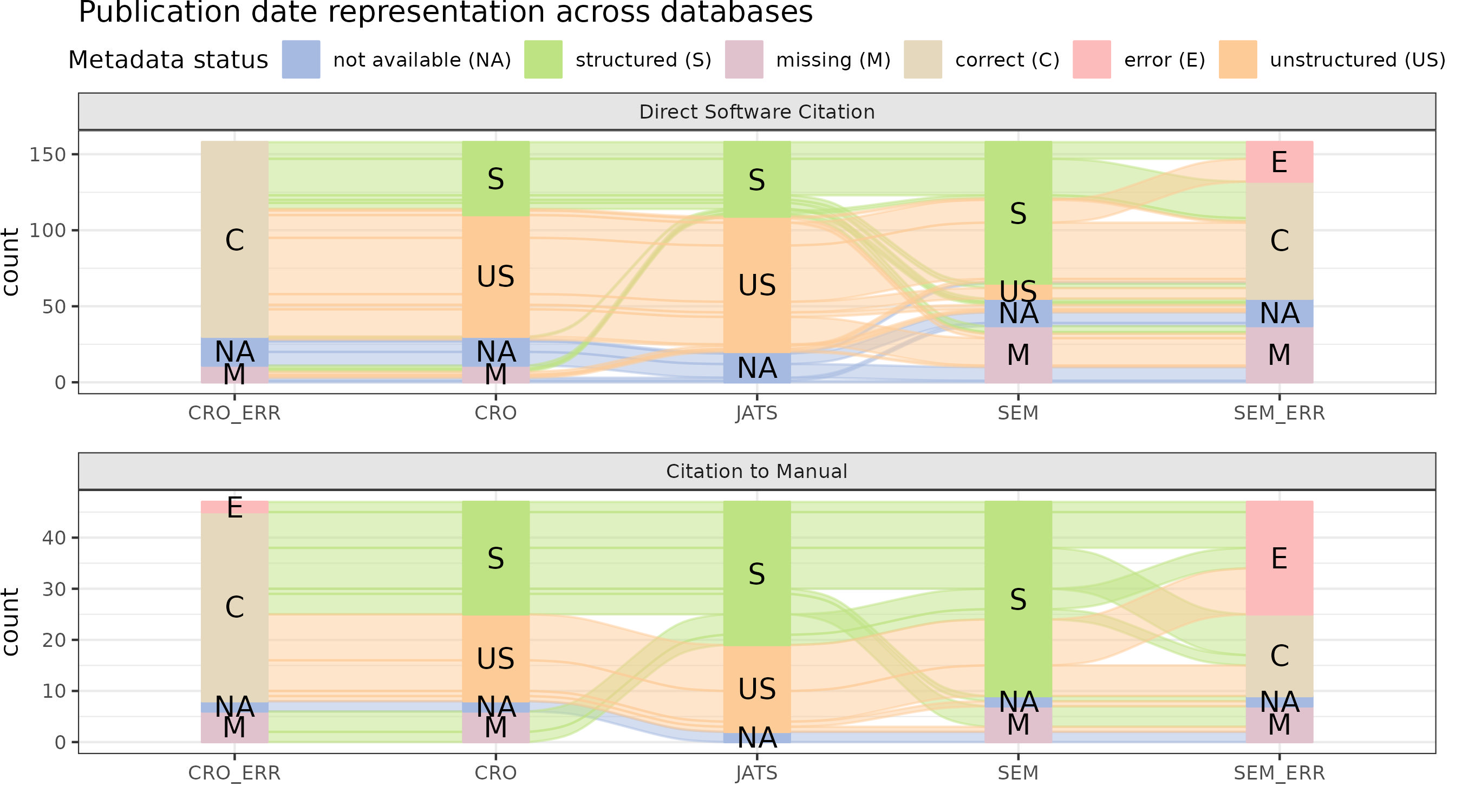}
    \caption{Alluvial plot illustrating the availability, structure, and correctness of software release date representations within \emph{Direct Software Citations} and \emph{Manuals} following the principle of Figure~\ref{fig:plot_ex}.}
    \label{fig:db_year}
\end{figure*}
The publication date is commonly included by publishers in both \emph{Direct Citation} (87.3\%) and \emph{Manual} citations (95.7\%). 
It is often represented in a structured manner for \emph{Manuals} (59.6\%) but less frequently for \emph{Direct Citations} (31\%). 
Crossref and Semantic Scholar only loose information on publication date in few references with 1.4\% and 7.4\% for \emph{Direct} and 0\% and 5\% for \emph{Manual}. 
In turn, information is added by Semantic Scholar in 0.8\% of \emph{Direct} citations. 
Semantic Scholar manages to strongly increase the ratio of structured information, for both \emph{Direct} (76.9\%) and \emph{Manual} citations (95\%), while Crossref mainly reflects the structure of the publisher. 
Notably, Semantic Scholar retains structure in all cases except for 2.5\% \emph{Direct Citations}, and adds structure for 45.5\% of \emph{Direct Citations} and 37.5\% of \emph{Manuals}, while Crossref adds structure in 2.7\% of \emph{Direct Citations}.
High numbers of errors are present in Semantic Scholar for \emph{Direct Citations} (25.2\%) and \emph{Manuals} (57.9\%). 
For Crossref, few errors are present in \emph{Manuals} (5.1\%), all due to incomplete information.
Errors in Semantic Scholar for \emph{Direct Citation} are distributed between wrong information (88.5\%), incomplete information (7.7\%), and misrepresentation (3.8\%), and for \emph{Manuals} between wrong information (90.9\%) and incomplete information (9.1\%).

\newcommand{\rot}[1]{\rotatebox{90}{#1}}
\newcolumntype{P}[1]{>{\centering\arraybackslash}p{#1}}
\newcolumntype{L}[1]{>{\raggedright\arraybackslash}p{#1}}
\newcolumntype{R}[1]{>{\raggedleft\arraybackslash}p{#1}}
\newcolumntype{C}{>{\centering\arraybackslash}X}

\begin{table}[!h]
    \centering
\resizebox{0.85\columnwidth}{!}{%
\begin{tabularx}{\textwidth}{XXP{.8cm}|R{.8cm}|R{.7cm}R{.7cm}R{1cm}R{.7cm}R{1cm}R{.7cm}|R{.8cm}R{.7cm}R{.8cm}R{.7cm}}
    \toprule
\multirow{3}{*}{\rot{Metadata}} & \multirow{3}{*}{\rot{Citation}} & \multirow{3}{*}{\rot{Database}}  &  & \multicolumn{6}{c}{Structure} & \multicolumn{4}{|c}{Correctness} \\ 
&&&&&&&&&&&&\\
 & & & n & NA & (\%) & US & (\%) & S & (\%) & C & (\%) & E & (\%) \\ 
  \midrule
\multirow{6}{*}{\rotatebox{90}{Name}} & D & JATS & 158 &   9 & 5.7 & 129 & 81.6 &  20 & 12.7 & - & - & - &-  \\ 
   & D & CRO & 147 &  11 & 7.5 & 120 & 81.6 &  16 & 10.9 & 136 & 92.5 & 0 & 0 \\ 
   & D & SEM & 121 &   6 & 5.0 &  86 & 71.1 &  29 & 24.0 & 111 & 91.7 &   4 & 3.3 \\ 
   & M & JATS &  47 &   4 & 8.5 &  29 & 61.7 &  14 & 29.8 & - & - &  -& - \\ 
   & M & CRO &  41 &   7 & 17.1 &  24 & 58.5 &  10 & 24.4 &  34 & 82.9 &  0& 0  \\ 
   & M & SEM &  40 &   3 & 7.5 &  13 & 32.5 &  24 & 60.0 &  35 & 87.5 &   2 & 5.0 \\\midrule
  \multirow{6}{*}{\rotatebox{90}{Identifier}} & D & JATS & 158 &  86 & 54.4 & 0 & 0 &  72 & 45.6 &  -&  -&  -& - \\ 
   & D & CRO & 147 &  90 & 61.2 &   6 & 4.1 &  51 & 34.7 &  57 & 38.8 & 0 & 0  \\ 
   & D & SEM & 121 & 103 & 85.1 &  13 & 10.7 &   5 & 4.1 &  15 & 12.4 &   3 & 2.5 \\ 
   & M & JATS &  47 &  28 & 59.6 &  0& 0 &  19 & 40.4 & - & - & - & - \\ 
   & M & CRO &  41 &  29 & 70.7 &   1 & 2.4 &  11 & 26.8 &  12 & 29.3 & 0 & 0 \\ 
   & M & SEM &  40 &  40 & 100.0 & - & - & - & - & - & - & - &-  \\\midrule
  \multirow{6}{*}{\rotatebox{90}{Creator}} & D & JATS & 158 &  18 & 11.4 &  88 & 55.7 &  52 & 32.9 & - & - & - & - \\ 
   & D & CRO & 147 &  36 & 24.5 &  87 & 59.2 &  24 & 16.3 &  94 & 63.9 &  17 & 11.6 \\ 
   & D & SEM & 121 &  41 & 33.9 &  39 & 32.2 &  41 & 33.9 &  55 & 45.5 &  25 & 20.7 \\ 
   & M & JATS &  47 &   1 & 2.1 &  19 & 40.4 &  27 & 57.4 & - & - & - &-  \\ 
   & M & CRO &  41 &  16 & 39.0 &  16 & 39.0 &   9 & 22.0 &  17 & 41.5 &   8 & 19.5 \\ 
   & M & SEM &  40 &   6 & 15.0 &   7 & 17.5 &  27 & 67.5 &  29 & 72.5 &   5 & 12.5 \\\midrule
  \multirow{6}{*}{\rotatebox{90}{Version}} & D & JATS & 158 &  60 & 38.0 &  92 & 58.2 &   6 & 3.8 & - & - & - & - \\ 
   & D & CRO & 147 &  64 & 43.5 &  80 & 54.4 &   3 & 2.0 &  81 & 55.1 &   2 & 1.4 \\ 
   & D & SEM & 121 &  53 & 43.8 &  67 & 55.4 &   1 & 0.8 &  51 & 42.1 &  17 & 14.0 \\ 
   & M & JATS &  47 &  36 & 76.6 &  10 & 21.3 &   1 & 2.1 & - & - & - & - \\ 
   & M & CRO &  41 &  32 & 78.0 &   9 & 22.0 & 0 & 0 &   9 & 22.0 & 0 & 0 \\
   & M & SEM &  40 &  31 & 77.5 &   8 & 20.0 &   1 & 2.5 &   6 & 15.0 &   3 & 7.50 \\\midrule
  \multirow{6}{*}{\rotatebox{90}{Date}} & D & JATS & 158 &  20 & 12.7 &  89 & 56.3 &  49 & 31.0 & - & - & - &-  \\ 
   & D & CRO & 147 &  19 & 12.9 &  80 & 54.4 &  48 & 32.7 & 128 & 87.1 & 0 & 0 \\ 
   & D & SEM & 121 &  18 & 14.9 &  10 & 8.3 &  93 & 76.9 &  77 & 63.6 &  26 & 21.5 \\ 
   & M & JATS &  47 &   2 & 4.3 &  17 & 36.2 &  28 & 59.6 & - & - & - & - \\ 
   & M & CRO &  41 &   2 & 4.9 &  17 & 41.5 &  22 & 53.7 &  37 & 90.2 &   2 & 4.9 \\ 
   & M & SEM &  40 &   2 & 5.0 & 0 & 0 &  38 & 95.0 &  16 & 40.0 &  22 & 55.0 \\ 
   \midrule
\end{tabularx}
   }
   \caption{Overview of \emph{Direct Software Citation} and \emph{Manual} representation by publisher, Crossref, and Semantic Scholar across different meta-data. Column ``Citation'', distinguishes between \emph{Direct Citations} (D) and \emph{Manuals} (M); ``Database'' distinguishes the data source between publisher (JATS), Crossref (CRO), and Semantic Scholar (SEM); ``Structure'' NA refers to not available information, US to unstructured information, and S to structured information; ``Correctness'' C refers to correct information and E to wrong information. The JATS data has no correctness information because it is correct by definition. The values of NA, US, and S sum to one, and the values of NA, C, and E, also, because a correctness cannot be determined for not represented information. Missing references are not included in any counts. }
   \label{tab:res_summary}
\end{table}

\subsubsection{Description and Type of Citation}
Description and type of citation are considered less crucial to software citation as the information discussed so far. 
Therefore, the results are only briefly discussed with the corresponding alluvial plots available in supplementary Figures S10 and S11. 
A software description is included by publishers in about half of \emph{Direct Citations} (49.4\%) and \emph{Manual} citations (57.4\%), with only some information represented in a structured manner with 8.2\% of \emph{Direct Citations} and 31.9\% of \emph{Manuals}. 
Crossref and Semantic Scholar only loose information on software descriptions in rare cases for \emph{Manuals} with 2.40\% and 5\%, respectively. 
Semantic Scholar manages to increase the ratio of structured information, for both \emph{Direct Citations} (18.2\%) and \emph{Manual} citations (55\%), while Crossref directly reflects publisher structure, when information is not lost. 
Errors are rare for description and only appear in 3\% of \emph{Direct Citation} in Semantic Scholar and 1.3\% in Crossref.

The type of citation is commonly not included in both \emph{Direct Citation} (26.6\%) and \emph{Manual} citations (34\%), and never represented in a structured manner. 
Information is in some cases lost for \emph{Direct Citations} (Semantic Scholar 9.1\%, Crossref 3.4\%) and \emph{Manuals} (Semantic Scholar 2.5\%, Crossref 7.3\%). 
Structure is mainly represented as by the publisher for both databases, with Semantic Scholar adding some structure to \emph{Direct Citations} (0.8\%). 
Errors only appear for \emph{Direct Citations} with 15.8\% in Semantic Scholar 2.9\% in Crossref.

\section{Limitations}

In general, manual annotation was required to assess the quality of formal software citation, which is associated with a high manual effort.
To make this annotation feasible, SoMeSci was chosen as a basis because it allows the extension of existing annotations, strongly reducing the required effort.  
Overall, we consider the sample size sufficient to make reliable statements about the quality of formal software citation and its representation in bibliographic databases.
Our assessment is that it would be possible to perform further large-scale analyses examining the completeness of \emph{Direct Citations}, however, we assess an automatic evaluation of database accuracy as extremely challenging based on the problems we faced during the annotation.

Overall, we consider the data selection of SoMeSci as suited for the given analysis because it includes articles using software and articles creating software, covering formal citations from both groups.
But there are also limitations resulting from this data selection. 
A main drawback is that a large amount of articles is published by PLoS, which leads to a bias in publishers, affecting specifically the analyses on database accuracy. 
We argue that PLoS is a representative choice regarding the handling of software citation because PLoS has a high interest in software. 
PLoS ONE, the largest journal published by PLoS with a considerable margin, allows software submission and publishes corresponding software articles, but also encourages proper software mention, with the journal policy stating that authors should provide all software with versions and related references that are used for statistical analyses\footnote{\url{https://journals.plos.org/plosone/s/submission-guidelines}, accessed February 23th, 2024}. 
Moreover, an analyses excluding PLoS articles showed the same general trends with the same major issues from both the publisher side and the bibliographic databases.
While these findings are based on a small sample size, they highlight that the identified issues of formal software citation representation are not specific to one publisher but exist broadly across the current bibliographic infrastructure. 
However, it should be noted that specific publishers might already handle software citation in a suited manner.

Furthermore, all articles are available from the PubMed Central Open Access set and, therefore, have a selection bias towards life sciences and open access publications. 
This bias is likely to influence the type of software used within articles and can also have an influence on how the corresponding software is published and cited, e.g., a differing ratio in commercial and open access software and biomedical tools as compared to other disciplines. 
It could, for instance, be argued that the use of archive links is more common in the domain of computer science or other software heavy domains where the reuse and adaption of source code is more common.  

The article selection for the analyses spans a range of publication dates from 2007 and 2020 and, can, therefore, only make statements about software citation in this time span, and the amount of available data is not sufficient to analyse trends in formal software citation throughout this time. 
However, a previous large-scale analysis including data up to 2021 has shown that the overall number of software citations stayed at a plateau since 2009~\citep{Schindler2022TheRO}.
Based on these findings we assume that there were no major changes in software citation practices in the examined time frame and our results are valid.
A benefit of the given article selection is that bibliographic data providers had sufficient time to react and represent the formal software citations given within the articles, making the examined dataset well suited to assess the representation of formal software citation in bibliographic databases. 

\section{Discussion and Conclusion}

We analyzed data quality of software citations regarding the three quality dimensions structure, completeness and accuracy and found significant issues across all stages of the data life cycle.
Starting with the references as given by the authors, the analyses of software citation types showed a strong trend towards citation of \emph{Software Articles}, which is suited for identifying a software and its developer, but does not allow an identification of the code base. 
A reason why this practice might be well adapted is that authors are familiar with article citation.
This means that the majority of formal software references does not enable the identification of the used code base. 
Furthermore, we showed that authors using software articles also provide significantly less information in article full-text documents to enable the identification of software code base. 
The second largest groups are \emph{Direct Citation}, which---properly executed---is the most complete way to cite software and is discussed in detail below, and we identified a small trend for the citation of software manuals. 
Compared to prior studies, we observe a higher amount of \emph{Direct Citations} but confirm the results that \emph{Software Articles} are the most cited resource for scientific software. 
This shows that better awareness on the author side is required since the majority of used software references are unsuited to represent software, with the majority resulting from the outdated notion that articles are more valuable scientific contributions than software~\citep{hafer2009assessing}. 

In practice, the type of citation is influenced by the citation recommendation made by the developers~\citep{Du2022}, often placed on the software download website or given with the newly established software citation format\footnote{\url{https://citation-file-format.github.io/}} (CFF).
Authors are likely to follow this recommendations in order to provide the desired attribution to the software developers, and because the provided information can be readily used. 
However, these recommendations often do not recommend \emph{Direct Citations} but other citations forms such as \emph{Manuals} and, therefore, omit essential information such as the version that identify the code base and are part of the research provenance.  
For the widely used statistical framework R, for instance, the recommended citation style is a citation of the  \emph{Manual}, which does not include a version number.  
Further, authors that publish software articles have a vested interest in the article being cited and are likely to recommend the citation of the \emph{Software Article} because they do otherwise not receive attribution and impact for the creation of the software. 
This means that action is also required by the developers of software to update software citation recommendations so they do not impose a conflict on authors. 

The analysis of citation completeness showed that software is almost always identifiable from \emph{Direct Citations} in 88\%--99\% of cases, even so the practice is only used in about 23\% of formal software references.
However, unique identifiers and archive links have almost never been utilized to identify software, showing that this is not a common practice, yet.
Attribution of developers is also possible for a majority of references in 88\% of cases, while the information regarding code base is only identifiable in 66\%--73\%.
In general, the use of version numbers was found to be a common practice, while  release dates are almost never used, and mostly in cases where software developers utilize release dates, e.g., ``Matlab r2023a''. 
Overall, the completeness of \emph{Direct Citations} is at a good level, while the situation regarding \emph{Manual} citations is similar, with software identification and attribution being given is most cases, while code base identification is at a notably lower level (see Figure~\ref{fig:comp_id}).
Overall, the citation completeness could be further improved with better awareness regarding the importance of code base identification for reproducibility and research provenance, which further highlights the argument made above.  

Information provided by publishers is mostly unstructured for \emph{Direct Citations}, hindering systematic analyses of software usage in science, which rely on structured information.
Creator and date information are structured to some extend, but still less than 50\% of cases, while the name is structured in even less cases and versions only in single instances. 
The only information that is almost always consistently structured are identifiers. 
In general, information from \emph{Manuals} is more often structured than information from \emph{Direct Citations} (see Table \ref{tab:res_summary}), which reflects the similarity of \emph{Manuals} to scholarly articles in contrast to other research objects such as software and data.

Bibliographic databases take information about references either directly from the publishers or by analysing the reference lists from scholarly articles.
Crossref was found to mostly take over information from the publishers retaining the structure and information. 
Especially, names and identifiers are almost identical in their representation, while some information for versions, creator, and identifier is lost.
With respect to completeness, Crossref systematically represents author names by the lastname of the first author for all references, but maintains all information for the actual elements.
To access the data, a persistent identifier, i.e., DOI can be used to link the reference of a citing article to the actual entry in Crossref.
However, this raises a problem, when such an identifier is not present as in the case of most software citations.
Moreover, Crossref omits a small amount of direct software citations and manuals, as compared to software articles.
Overall, Crossref is not performing any special treatment for software citation, but is mostly successful in representing the information available from publishers. 

With 22.9\%, Semantic Scholar omits a significant amount of direct software references.
It employs an automatic approach to link similar references to the same element.
While this increases the structure of information to some extend, it introduces errors when it comes to direct software citations and software manuals. 
Wrong information (14\%) as well as duplicated information (18\%) is present in 30\% of represented references in Semantic Scholar, which likely results from adding information that was found by erroneous linking of different references to the same element (see Listing \ref{list:database_error}).
Moreover, Semantic Scholar looses information within the software citations it retains. 
It does, for instance, drop a high number of versions, and the majority of URLs. 
It also introduces a high number of errors in versions, creators, and dates.
Overall, software citation representation in Semantic Scholar is poor, because it seems that the underlying implementation has not been adapted to handle \emph{Direct Software Citations}. 
Specifically the concepts of versions and URLs that are common in software citations are not represented in Semantic Scholar. 
However, it succeeds in improving some references, and it can be assumed that with proper adaption it could be successful in representing and even adding structure to software citations, when the original published information lacks it. 

In general, both databases are currently unable to adequately represent software citations, as proper handling of the \emph{Direct Citations} does not seem to be implemented in either. 
Some fault does also lie at the publisher site, where there is also no suited format for direct software citations, and instead software citations are adapted to match the fields used for regular citations. 
Based on the current systems, systematic analyses of formal software citations in scientific articles are not possible. 
In general, our results show that both, publishers and bibliographic databases, need to update their infrastructure to create suited and machine-readable software representations, as suggested by \citet{stall2023journal}. 
Therefore, we urge the providers of bibliographic data to update their implementations to take the intricacies of software citation into account, to enable and facilitate systematic representation and analyses of formal software citation in the future. 
We recommend to include at least two different views to software: (1) provide all information about the particular software as given by the authors to enable reproducibility, and (2) link different versions of the same software to a common element to credit creators of such.
A spot check in the Scopus database, which is generally known as a high quality data source for bibliometric studies~\citep{Baas2020} revealed similar issues, for instance, unstructured data, missing specialized treatment for software with different version, or citations to the same software that are not linked and consequently evaluated independently.
The data published in the scope of this work can serve as a starting point for analyzing formal software citations and the requirements for representing them and can serve as initial training for machine learning methods.

Authors looking to cite software in their articles face a conflict. 
Since bibliographic databases currently fail at representing software citations, it cannot be recommended to use direct software citations in scientific publications, only. 
Instead, software usage should additionally be indicated by mentioning the software in the full-text document with all information required to identify it and citing corresponding software articles.
This practice allows systematic analyses by employing methods such as the SoftwareKG information extraction pipeline~\citep{Schindler2022TheRO}, but also gives direct credit to developers. 
There is still a strong argument for the use of formal software citation, as it clearly identifies software and provides credit without requiring elaborate machine learning methods to extract the knowledge. 
However, as long as providers of bibliographic data do not adequately represent direct software citations this knowledge stays inaccessible. 
We hope this situation can quickly be resolved by updates to existing bibliographic databases. 
In this context, we advocate the use of formal software citations, as further adaption of this practice increases the need and urgency to address this problem.

\paragraph*{Software}
In the following all software used during this investigation is listed including software citations and software articles for all software for which they exist. 
We used both Python~\citep{python} 3.8.16 and R~\citep{R} 4.3.0 for data processing. 
For Python we further used the package articlenizer R-14.06.2021~\citep{articlenizer}. 
For R we used the packages tidyverse~\citep{tidyverse,tidyverse2} 2.0.0 and magrittr~\citep{magrittr} 2.0.3 for data processing, patchwork~\citep{patchwork} 1.1.2, ggalluvial~\citep{ggalluvial} 0.12.5, easyalluvial~\citep{easyalluvial} 0.3.1, and xtable~\citep{xtable} 1.8-4 for output generation, and DescTools~\citep{desctools} 0.99.48 and rcompanion~\citep{rcompanion} 2.4.30 for statistical analysis. 
Further, we used RStudio~\citep{rstudio} 2023.3.1.446 for development and Quarto~\citep{quarto} 1.2.475 to generate a literate data analysis document. 

\paragraph*{Data and materials availability} 
The script and data to replicate the analyses described in this work are available at Zenodo~\citep{somescicitation} and Github\footnote{\url{https://github.com/dave-s477/SoMeSci_Citation}}, with the data being made available together with the original {\sc SoMeSci} data~\citep{somesci_data} at Github\footnote{\url{https://github.com/dave-s477/SoMeSci}} and Zenodo\footnote{\url{https://doi.org/10.5281/zenodo.4968738}}.

\paragraph*{Funding}
This work was supported by the Deutsche Forschungsgemeinschaft (DFG, German Research Foundation) SFB 1270/2: 299150580.

\paragraph*{Author Contributions}
Conceptualization, D.S., T.H., S.S., F.K.; 
Data curation, D.S., T.H., F.K.; 
Formal Analysis, D.S., F.K.; 
Funding acquisition, S.S., F.K.; 
Investigation, D.S., T.H.; 
Methodology, D.S., T.H., F.K.; 
Project administration, S.S., F.K.; 
Resources, D.S., T.H.; 
Software, D.S., T.H.; 
Supervision, F.K.; 
Validation, D.S., F.K.; 
Visualization, D.S., F.K.; 
Writing---original draft, D.S., T.H., S.S., F.K.; 
Writing---review \& editing, D.S., F.K.;

\paragraph*{Competing interests}
The authors declare no competing interests.

%
%
\bibliographystyle{apacite}
\bibliography{references}

\end{document}